\documentstyle[epsf,12pt]{article}
\newcommand{\ewxy}[2]{\setlength{\epsfxsize}{#2}\epsfbox[10 60 640 570]{#1}}
\newcommand{\ltaeq}{$\raisebox{-.6ex}{$\stackrel{\textstyle{<}}{\sim}$}$}
\newcommand{\gtaeq}{$\raisebox{-.6ex}{$\stackrel{\textstyle{>}}{\sim}$}$}
\newcommand{\bra}[1]{\left< #1 \right|}

\newcommand{\ket}[1]{\left| #1 \right>}\begin{document}
\newcommand{\dla}{\stackrel{\leftarrow}{D}}
\newcommand{\vla}{\stackrel{\leftarrow}{v}}
\newcommand{\dra}{\stackrel{\rightarrow}{D}}
\newcommand{\dslash}{\! \not \!\! D}
\newcommand{\vslash}{\! \not \!\! v}
\begin{titlepage}
\begin{flushright}
{\small 
FSU-SCRI-96-43\\
GUTPA/96/4/1\\
OHSTPY-HEP-T-96-009}
\end{flushright}

\begin{center}
{\Large $B$ Decay Constants from NRQCD with Dynamical Fermions}\\[5mm]

{\bf S.~Collins\footnote{Present address University of Glasgow, Glasgow, Scotland G12 8QQ}, U.~M.~Heller, and J.~H.~Sloan}\\
SCRI, Florida State University, Tallahassee, Fl 32306-4052, USA\\[3mm]

{\bf J.~Shigemitsu}\\
The Ohio State University, Columbus, OH 43210, USA\\[3mm]

{\bf A.~Ali~Khan\footnote{UKQCD Collaboration} and C.~T.~H.~Davies${\vphantom{\Large A}}^2$}\\
University of Glasgow, Glasgow, G12 8QQ, Scotland 

\end{center}

\begin{abstract}
We present a lattice investigation of the heavy-light meson decay
constants using Wilson light quarks and NRQCD heavy quarks, partially
including the effects of dynamical sea quarks. We calculate the
pseudoscalar and vector decay constants over a wide range in heavy
quark mass and are able to perform a detailed analysis of heavy quark
symmetry. We find consistency between the extrapolation of the NRQCD
results and the static case, as expected. We find the slope of the
decay constants with $1/M$ is significantly larger than naive
expectations and the results of previous lattice calculations.  For
the first time we extract the non-perturbative coefficients of the
slope arising from the $O(1/M)$ heavy quark interactions separately
and show the kinetic energy of the heavy quark is dominant and
responsible for the large slope.  In addition, we find that
significant systematic errors remain in the decay constant extracted
around the $B$ meson mass due to truncating the NRQCD series at
$O(1/M)$.  We estimate the higher order contributions to $f_B$ are
approximately $20\%$; roughly the same size as the systematic errors
introduced by using the Wilson action for light quarks.
\end{abstract}
PACS numbers: 12.38.Gc, 12.39.Hg, 13.20.He, 14.20.Nd
\thispagestyle{empty}
\end{titlepage}

\section{Introduction}
\label{intro}
The decay constant of the $B$ meson, $f_B$, is of fundamental importance
to tests of the Standard Model because it is needed in the
determination of the CKM matrix.
While one of the simplest weak matrix elements to
study, it is not very well known and dominates the uncertainty in
extracting $V_{td}$, one of the smallest CKM elements, from
$B-\bar{B}$ mixing. The best chance of an experimental measurement of
$f_B$ lies in the decay $B^+\rightarrow
\tau^+ + \nu$.  However, this decay has yet to be observed, and $f_B$
appears in combination with $V_{ub}$, which is only known to within a
factor of 2.

Theoretical calculations using for example QCD sum rules within the
framework of Heavy Quark Effective Theory~(HQET)~\cite{isgur,neublong}
and lattice field theory have been actively pursued over the last few
years with the current world average standing at
$f_B=200\pm20\%$~MeV~\cite{chrisa,qcdsum}.  As a first principles
approach, lattice simulations offer ultimately the most reliable
calculation of $f_B$. The initial problem of simulating a $b$ quark
with Compton wavelength of the order of the typical lattice spacing
presently achievable was first solved by using the static
approximation~\cite{eichtinitial,eichten}, where the heavy quark is
treated as a static colour source within the heavy-light
meson. Simulations using conventional relativistic methods around the
$D$ meson showed the corrections to the static approximation in this
region, and hence also at the $B$ meson, are much larger than
naively expected. This has contributed to the sizable uncertainty
remaining in $f_B$, shown above.  The determination of $f_B$ directly
at $M_B$ on the lattice is required, and this is now possible with the
development of Non Relativistic
QCD~(NRQCD)~\cite{bethlep,biglep}. Using this approach the error in
the decay constant can be significantly reduced with the computational
power currently available. Initial calculations have already been
performed by Hashimoto~\cite{hash} and Davies~\cite{cdavies}.

This analysis forms the second stage of a project applying NRQCD to
heavy-light systems and studying heavy quark symmetry at the $B$ meson.
Our previous papers have dealt with the $B$ meson spectrum in both the
quenched approximation~\cite{arifaspec} and partially including the
effects of sea quarks \cite{spec}. We now present our results for the
decay constants of the heavy-light mesons from dynamical
configurations. Note that the combination $f_B^2B_B$, where $B_B$ is
the mixing parameter for the $B$ system, is the experimentally
relevant quantity, and thus a calculation of $B_B$ is also needed.

The paper is organised as follows. In section~\ref{simdet} we describe
the simulation details including the systematic errors associated with
the calculation. The meson operators required to compute the decay
constant to $O(1/M)$ are set out in section~\ref{mesop} and we discuss
computing the decay constant consistently to this order in terms of
the evolution equation used. Next, we examine the predictions of heavy
quark symmetry for the behaviour of the decay constant in the heavy
quark mass limit. The contributions from the interactions of the heavy
quark at $O(1/M)$ are analysed and combinations of decay constants
which isolate these individual contributions are
discussed. Section~\ref{results} illustrates our results and fitting
analysis with a subset of the data. The heavy quark mass dependence of
the decay constants are investigated in section~\ref{hq_decay}; the
individual non-perturbative $O(1/M)$ coefficients of the slope of the
decay constant are extracted. Finally we present our conclusions in
section~\ref{conc}.

\section{Simulation Details}
\label{simdet}
The simulations were performed using 100 $16^3\times32$ gauge
configurations at $\beta=5.6$ with two flavours of staggered dynamical
sea quarks with a bare quark mass of $am_{sea}=0.01$. These
configurations were generously made available by the \mbox{HEMCGC}
collaboration; more details can be found in~\cite{hemcgc}. We fixed
the configurations to the Coulomb gauge. The light quark propagators
were generated using the Wilson fermion action, without an $O(a)$
improvement term, at two values of the hopping parameter,
$\kappa=0.1585$ and $0.1600$.  The former corresponds to a quark mass
close to strange, where $\kappa_s=0.1577$ from $M_\phi$, while
$0.1600$ is somewhat lighter, with $\kappa_c=0.1610$. 

With only two values of light quark mass it is not possible to perform
a trustworthy chiral extrapolation. In addition, the systematic errors
associated with the light quarks, discussed below, affect the
determination of the bare strange quark mass; $\kappa_s$ extracted
from pseudoscalar mesons, using the lowest order chiral mass
dependence to find the `experimental mass' of the pure $s\bar{s}$
pseudoscalar, differs from $\kappa_s$ obtained from the vector meson
$\phi$. Thus, most of the results will be presented for
$\kappa_l=0.1585$.

In this simulation we truncate the NRQCD series at $O(1/M_0)$, where
$M_0$ is the bare heavy quark mass, and the
action takes the form:
\begin{equation}
S = Q^{\dagger}(D_t + H_0 + \delta H) Q
\end{equation}
where 
\begin{equation}
H_0 = -\frac{\Delta^{(2)}}{2M_0}\hspace{0.5cm} \mbox{and} 
\hspace{0.5cm} \delta H = -c_B\frac{\sigma\cdot B}{2M_0}.
\end{equation}
Tadpole improvement of the gauge links is used throughout i.e. they
are divided by $u_0$, where we use $u_0=0.867$ measured from the
plaquette, and the hyperfine coefficient is given the tree-level value
$c_B=1$. We use the standard Clover-leaf operator for the B field. The
heavy quark propagators were computed using the evolution equation~\cite{nrqcdups}:
\begin{eqnarray}
G_1 & = & \left( 1-\frac{aH_0}{2n}\right)^n 
U_4^{\dagger} \left(1-\frac{aH_0}{2n}\right)^n 
\delta_{\vec{x},0}\nonumber\\
G_{t+1} & = & \left(1-\frac{aH_0}{2n}\right)^n 
U_4^{\dagger} \left(1-\frac{aH_0}{2n}\right)^n 
(1-a\delta H)G_t \hspace{0.5cm} (t>1)
\label{evol}
\end{eqnarray}
where $n$ is the stabilising parameter. In the static limit this
reduces to
\begin{equation}
G_{t+1} - U_4^\dagger G_t = \delta_{x,0}.
\end{equation}

We generated
heavy quark propagators at 11 values of ($aM_0$,n) corresponding to
(0.8,5), (1.0,4), (1.2,3), (1.7,2), (2.0,2), (2.5,2), (3.0,2),
(3.5,1), (4.0,1), (7.0,1) and (10.0,1), and the static limit. This
roughly corresponds to a range of meson masses from $M_B/3$ to $4M_B$
and is sufficient for a reasonable investigation of heavy quark
symmetry. However, it is not possible to simulate the $D$ meson on
this lattice using NRQCD; a larger lattice spacing is required,
$\beta^{n_f=2}\ltaeq 5.5$ or $\beta^{n_f=0}\ltaeq 5.85$.  

Details of the construction of meson operators, the smearing functions
used and our fitting analysis can be found in~\cite{spec}.  As
discussed in this reference there are large systematic errors
associated with this simulation arising from the use of Wilson light
fermions. The emphasis of our analysis will be on determining the
heavy quark mass dependence of the decay constants and extracting the
nonperturbative coefficients of the heavy quark expansion to
$O(1/M)$. The systematic error arising from the light quark in these
$O(\Lambda_{QCD}^2)$ coefficients is naively estimated to be a
relative error of $O(\Lambda_{QCD}a)\sim 20\%$ for this
ensemble~(where we take $a\Lambda_{QCD}=a\Lambda_V=0.185$ for these
configurations).  The truncation of the NRQCD series and the use of
the tree-level value for the hyperfine coefficient introduce
absolute errors of $O(\Lambda_{QCD}(\Lambda_{QCD}/M)^2)$ and
$O(\alpha_S\Lambda_{QCD}(\Lambda_{QCD}/M))$ respectively.  Both
correspond to approximately $1\%$ errors in the value and
$10\%$ errors in the slope of the decay constant at $M_B$. The coefficients of
the quadratic term in the heavy quark expansion cannot be correctly
determined in this analysis. The size of the systematic errors induced by the
light quark action are indicated by the large uncertainty in the
lattice spacing; we use $a^{-1}=1.8{-}2.4$~GeV. This gives
$aM_0=2.4{-}1.7$ as corresponding to the bare $b$ quark mass. We are
currently repeating our analysis using tadpole-improved Clover light
fermions~\cite{clovernrqcd}.

\section{Meson Operators}
\label{mesop}
The pseudoscalar decay constant is defined by
\begin{eqnarray}
\bra{0} A_0 \ket{PS} & = & f_{PS}M_{PS},
\end{eqnarray}
where $A_0=\bar{q}\gamma_5\gamma_0 h$ at tree level and $q$ and $h$
represent 4-component light and heavy quark fields respectively. In
this convention the experimental value of $f_\pi$ is $131$~MeV.
Similarly, the vector decay constant is given by
\begin{eqnarray}
\bra{0} V_i \ket{V_i} & = & \epsilon_{i} f_V M_V,\footnotemark
\end{eqnarray}
\footnotetext{This is a more convenient definition than the more commonly used $\bra{0} \bar{q}\gamma_i h\ket{V_i}  =  \epsilon_{i} M_V^2/f_V$.}
where $V_i=\bar{q}\gamma_i h$ at tree level.
The calculation of these matrix elements, and thus decay
constants, to $O(1/M)$ in NRQCD requires several operators.  At tree
level these can be obtained by relating $h$ of full QCD to the
2-component NRQCD fields, $Q$ and $\tilde{Q}$ for heavy quarks and
heavy antiquarks respectively, using the Foldy-Wouthuysen
transformation.  To $O(1/M)$,
\begin{equation}
h = e^{-iS^{(0)}}\left(\begin{array}{c}  Q\\ \tilde{Q}^{\dagger}\end{array}
\right) \simeq (1 - iS^{(0)})\left(\begin{array}{c} Q \\ \tilde{Q}^{\dagger}
\end{array}\right),
\hspace{5mm} S^{(0)} = -\frac{i\gamma\cdot\vec{D}}{2M_0}
\end{equation}
With the convention that our meson states be built out of a heavy quark 
and a light antiquark,  the $\tilde{Q}$ field will not contribute to
the above matrix elements  through $O(1/M)$, and we obtain
\begin{equation}
\bra{0} \bar{q} \Gamma^{(4)} h\ket{P}_{QCD} = -\bra{0} q^\dagger_{34} 
 \Gamma^{(2)}
 Q\ket{P}_{NRQCD} + \bra{0} q^\dagger_{12}  \Gamma^{(2)} 
\left(\frac{i\sigma\cdot \vec{D}}{2M_0}\right) Q \ket{P}_{NRQCD} 
\label{foldycurr}
\end{equation}
where $\Gamma^{(4)}=\gamma_5\gamma_0$ and $\Gamma^{(2)}=1\!\!1$ for
the axial current, and $\Gamma^{(4)}=\gamma_i$ and
$\Gamma^{(2)}=i\sigma_i$ for the vector current.
 $q_{12}$ and $q_{34}$ denote the
upper and lower two components of $q$ respectively.  Our Euclidean space 
$\gamma$ matrices are chosen as,
\begin{equation}
\gamma_0 = \left(\begin{array}{cc} 1\!\!1 & 0 \\ 0 & -1\!\!1 \end{array} \right),\hspace{0.5cm}
\vec{\gamma} = \left( \begin{array}{cc} 0 & i\vec{\sigma} \\ -i \vec{\sigma} & 0
\end{array} \right).
\end{equation}

Beyond tree level all operators with the same quantum numbers can mix
under renormalisation.  Thus at $\small{O(\alpha/M)}$ the basis of
operators for the axial current, is given by~\footnote{In
the limit of zero light quark mass.}
\begin{equation}
{\cal O}_1  =  -q^\dagger_{34} Q \hspace{5mm} {\cal O}_2 = q^\dagger_{12} 
 \frac{i\sigma\cdot \vec{D}}{2M_0}Q \hspace{5mm}
{\cal O}_3  =  q^\dagger_{12}  \frac{-i \dla \cdot \sigma}{2M_0} Q 
\end{equation}
Using translation invariance, the zero momentum currents involving
${\cal O}_2$ and ${\cal O}_3$ are identical on the lattice and
comparing with equation~\ref{foldycurr} it is sufficient to
calculate matrix elements of the tree level operators ${\cal O}_1$ and
${\cal O}_2$. Similarly, the basis of operators for the vector
current,
\begin{eqnarray}
{\cal O}_1 & = & -iq^\dagger_{34} \sigma_i Q \hspace{5mm} 
{\cal O}_2 = q^\dagger_{12} \sigma_i \frac{-\sigma \cdot \vec{D}}{2M_0}Q 
\hspace{5mm}
{\cal O}_3  =  q^\dagger_{12} \frac{\dla \cdot \sigma}{2M_0} \sigma_i Q \\
{\cal O}_4 & = & q^\dagger_{12} \frac{-\vec{D}_i}{2M_0} Q \hspace{5mm}
{\cal O}_5  =  q^\dagger_{12} \frac{\dla_i}{2M_0} Q
\end{eqnarray}
is reduced, for the numerical computation, to ${\cal O}_1$, ${\cal
O}_2$ and ${\cal O}_2{-}{\cal O}_4$, where $\sigma$ matrix properties have also
been used. We computed the pseudoscalar and vector matrix elements
corresponding to this minimal set of operators by inserting a smeared
interpolating field with the same quantum numbers as ${\cal O}_1$ at
the source and all combinations of local currents at the sink for each
meson.

The perturbative calculation of the renormalisation factors needed to
relate the matrix elements in NRQCD to the currents in full QCD has
not yet been completed~\cite{nrqcdrenorm}. Thus, we use the static
renormalisation factors~\cite{eichten} and are restricted to the
tree level $O(1/M)$ corrections; we omit ${\cal O}_4$ for the vector
meson.  Since this correction is of
$O(\alpha_S\Lambda_{QCD}(\Lambda_{QCD}/M))$ it is roughly the same
size as that induced by unknown perturbative matching corrections in
the coefficient $c_B$, which are not included in the analysis. The
renormalisation factors in NRQCD depend on the heavy quark mass and
hence using the static value introduces an $O(\alpha_S)\sim 20\%$
error into the $O(1/M)$ coefficients extracted from our results.

It is possible that the power divergences which appear in NRQCD may
lead to significant non-perturbative contributions to the
renormalisation factors~\cite{sach1}. No clear indication of
`renormalon' effects has been found in other quantities calculated
perturbatively in NRQCD~\cite{nrqcdups}, however, we intend to compute the
renormalisation factors non-perturbatively to check this.

Strictly, our analysis is not fully consistent to $O(1/M)$.  Note that
the evolution equation given in equation~\ref{evol} induces an
$O(a(\Lambda_{QCD}/M))$ systematic error in the currents.
This is because it does not apply the relativistic correction operator
$(1-\delta H)$ on the initial timeslice. Since all subsequent
timeslices include this operator the transfer matrix is consistent to
$O(1/M)$. Thus, it is only in the amplitude of the meson correlator,
i.e.  the decay constant, that is missing this
$O(a(\Lambda_{QCD}/M))$ contribution. A better but
computationally more costly evolution equation is given by
\begin{eqnarray}
G_{t+1} & = & (1-\frac{a\delta H}{2})\left(1-\frac{aH_0}{2n}\right)^n 
U_4^{\dagger} \left(1-\frac{aH_0}{2n}\right)^n 
(1-\frac{a\delta H}{2})G_t \hspace{0.3cm} 
\end{eqnarray}
for all timeslices. Since our use of the Wilson action for light
 quarks also induces errors of this order of magnitude, we used
 equation~\ref{evol} in this project. In a comparison of these
 evolution equations using Clover fermions for the light
 quarks~\cite{clovernrqcd} we found the insertion of $(1-\delta H)$ on
 the source timeslice introduces $\sim 4\%$ correction to $\bra{0}
 {\cal O}_1 \ket{PS}_{NRQCD}$ and $\sim 3\%$ correction to $\bra{0}
 {\cal O}_2 \ket{PS}_{NRQCD}$ around $M_b$. Considering the systematic
 errors inherent in this calculation and our statistical accuracy of
 $\sim 2\%$, the omission of this correction will not significantly
 affect our findings.

\section{Heavy Quark Symmetry}
\label{hqsym}
In the heavy quark limit the decay constant can be parameterised
in terms of an expansion in $1/M$:
\begin{equation}
f\sqrt{M} = (f\sqrt{M})^\infty\left( 1 + \frac{c_P}{M} + O\left(\frac{1}{M^2}\right)\right).\label{fsqrtmq}
\end{equation}
The coefficient $c_P$ is determined by nonperturbative contributions
arising from the hyperfine interaction~($G_{hyp}$) and the kinetic
energy of the heavy quark~($G_{kin}$), which appear in the NRQCD action,
and the corrections to the current arising from the Foldy-Wouthuysen transformation. From first order perturbation theory
in $1/M$ about the static limit:
\begin{equation}
c_P = G_{kin} + 2d_MG_{hyp} + d_MG_{corr}/6
\label{slope}
\end{equation}
where $d_M=3$ and $-1$ for pseudoscalar and vector mesons
respectively. The integer factors in this equation are introduced for
convenience when comparing with HQET. 
\begin{eqnarray}
\lefteqn{\bra{0} q^\dagger \Gamma^{(2)} Q \ket{P}_\infty G_{kin}  = }\hspace{2.4cm} \nonumber\\
& & \bra{0} \int dy T\{q^\dagger \Gamma^{(2)}Q(0),Q^\dagger(\frac{-\vec{D}^2}{2}) Q(y)\} \ket{P}_{\infty},\label{gkin}
\end{eqnarray}
\vspace{-0.8cm}
\begin{eqnarray}
\lefteqn{\bra{0} q^\dagger \Gamma^{(2)} Q \ket{P}_\infty 2d_MG_{hyp}  = } \hspace{3.2cm}\nonumber\\
& & \bra{0} \int dy T\{q^\dagger \Gamma^{(2)}Q(0),Q^\dagger (\frac{-c_B\sigma\cdot B}{2}) Q(y)\}\ket{P}_{\infty},\label{ghyp}
\end{eqnarray}
\vspace{-0.8cm}
\begin{eqnarray}
\bra{0} q^\dagger \Gamma^{(2)} Q \ket{P}_\infty d_MG_{corr}/6 & = & \bra{0} q^\dagger\Gamma^{(2)} \frac{\sigma\cdot \vec{D}}{2} Q \ket{P}_\infty,\hspace{3.8cm}\label{cooreqn}
\end{eqnarray}
where $\ket{P}_\infty$ represents the meson in the limit of infinite
heavy quark mass, and equations~\ref{gkin}-\ref{cooreqn} are tree
level expressions. Note that the kinetic and hyperfine terms
contribute to the decay constant through the correction to the meson
wavefunction.

Equation~\ref{slope} suggests that the individual contribution from
each $O(1/M)$ term can be obtained separately by taking appropriate
combinations of the pseudoscalar and vector decay
constants, $(f\sqrt{M})_{PS}$ and $(f\sqrt{M})_V$ respectively.
For clarity below, we denote 
\begin{eqnarray}
 (f\sqrt{M})_{PS}^{uncorr}  & = & \frac{1}{\sqrt{M_{PS}}}  \bra{0} {\cal O}_1^{PS} \ket{PS}_{NRQCD}\\
\delta(f\sqrt{M})_{PS} & = & \frac{1}{\sqrt{M_{PS}}} \bra{0} {\cal O}_2^{PS}
\ket{PS}_{NRQCD} \\
(f\sqrt{M})_{PS}^{tot} & = & (f\sqrt{M})_{PS}^{uncorr} + \delta(f\sqrt{M})_{PS}
\end{eqnarray}
with similar quantities defined for the vector particle by
replacing PS with V. We also denote the
spin-average of the decay constants, 
\begin{eqnarray}
(\overline{f\sqrt{M}}) = ((f\sqrt{M})_{PS}^{uncorr} + 3(f\sqrt{M})_V^{uncorr})/4.
\end{eqnarray}

\begin{table}
\begin{center}
\begin{tabular}{|c|c|c|}\hline
 & expansion & slope\\\hline
$(\overline{f\sqrt{M}})$ & $(f\sqrt{M})^\infty\left(1+c_P'/M\right)$ & 
$c_P'  =  G_{kin}\label{cp_kin}$\\\hline
$(f\sqrt{M})^{uncorr}_{PS}/(f\sqrt{M})^{uncorr}_V$ & $1 + c_P''/M$ & 
$c_P'' = 8 G_{hyp}$\\\hline
$(f\sqrt{M})^{tot}_{PS}/(f\sqrt{M})^{tot}_V$& $1 + c_P'''/M$ & 
$c_P''' = 8 G_{hyp}+ 2 G_{corr}/3$\\\hline
\end{tabular}
\caption{Expansions about the static limit to $O(1/M)$ of 
combinations of the pseudoscalar and vector decay constant, with and
without the current corrections.
\label{expan}}
\end{center}
\end{table}

Table~\ref{expan} displays our notations for the expansions in $1/M$ and the
corresponding coefficients of $(\overline{f\sqrt{M}})$ and the ratio
of the pseudoscalar to vector decay constants, with and without the
current correction.  It is easy to see from this that both $G_{kin}$ and
$G_{hyp}$ can be extracted in a simple way from $c_P'$ and $c_P''$
respectively; $G_{corr}$ must be obtained from the combination
$c_P'''-c_P''$. Since we compute $\bra{0} {\cal O}_2 \ket{P}$
separately, however, $G_{corr}$ can be more accurately found by examining the
current correction in the infinite heavy quark mass limit:
\begin{equation}
G_{corr} = \frac{ [2M_0\bra{0} {\cal O}_2 \ket{P}]_\infty}{\bra{0} {\cal O}_1 \ket{P}_\infty}= \lim_{M_0\rightarrow \infty} \frac{2M_0\delta(f\sqrt{M})_{PS}}{(f\sqrt{M})^{uncorr}_{PS}}.
\label{corrdef}
\end{equation}

HQET provides an analogous decomposition of these
coefficients~\cite{neubert}.  However, since HQET constructs an
effective theory in terms of the heavy quark pole mass, as opposed to
the bare quark mass in NRQCD, only physical combinations of $G$'s,
i.e. $c_P$, $c_P'''$ and $c_P'=G_{kin}$, can be compared. Note that while
these coefficients are expected to be of $O(\Lambda_{QCD})\sim
200{-}500$~MeV~(and thus the corrections to the static limit at $M_B$
$O(\Lambda_{QCD}/M_B)\sim10\%$); the unphysical quantities $G_{hyp}$
and $G_{corr}$ need not be of this order.

In HQET $G_{corr}$ is related to the meson binding energy,
$G_{corr}=-\bar{\Lambda}$~\cite{neubert}. This follows from
the observation
\begin{eqnarray}
\bra{0}i\partial_\alpha(\bar{q} \Gamma^{(4)} h) \ket{P}_{\infty}
& = & (M_M - M_Q) v_\alpha \bra{0} \bar{q} \Gamma^{(4)} h \ket{P}_{\infty},\\
& = & \bar{\Lambda} v_\alpha \bra{0} \bar{q} \Gamma^{(4)} h \ket{P}_{\infty},
\label{bdead}
\end{eqnarray}
where $v_\alpha$ is the velocity of the heavy quark, and using the equations
of motion of the heavy and light quark. Note that in NRQCD the
R.H.S. of equation~\ref{bdead} becomes $E_{sim}^\infty v_\alpha
\bra{0} \bar{q}
\Gamma^{(4)} h
\ket{P}_{\infty}$ and following the arguments of~\cite{neubert} the
analogous expression in NRQCD can be obtained:
\begin{equation}
G_{corr} = -E_{sim}^\infty.
\label{coresim}
\end{equation}
At finite heavy quark mass this relation is broken by terms of
$O(1/M)$. Similarly, on the lattice, discretisation
modifies~\ref{coresim} and the operators in equation~\ref{corrdef}
must be renormalised.
\section{Results}
\label{results}
\subsection{Fitting results for the decay constants}
\label{fitres}
To begin with we extract the decay constant without the correction to
the current, i.e. from $\bra{0} {\cal O}_1 \ket{P}$. We performed a single
exponential vector fit to $C(1,l_{{\cal O}_1})$ and $C(1,1)$ for the
pseudoscalar and vector mesons. $C(1,l_{{\cal O}_i})$ denotes a
correlation function constructed at the source from an interpolating
field with appropriate quantum numbers for the vector or pseudoscalar
state and the heavy quark smeared with a ground S-state smearing
function~(described in~\cite{spec}); at the sink the local current
${\cal O}_i$ is inserted. In the limit of large values of
$t$
\begin{eqnarray}
C(1,l_{{\cal O}_1}) & = & Z_{l_{{\cal O}_1}} Z_{1} e^{-E_{sim} t},\\
C(1,1) & = & Z_{1}^2 e^{-E_{sim} t}.
\end{eqnarray}
By taking a bootstrap ratio of the amplitudes of these correlators the
decay constant can be obtained:
\begin{equation}
\sqrt{2}Z_{l_{{\cal O}_1}} = \bra{0} {\cal O}_1 \ket{P}_{NRQCD}/\sqrt{M} = (f\sqrt{M})^{uncorr}.
\end{equation}

\begin{figure}
\centerline{\ewxy{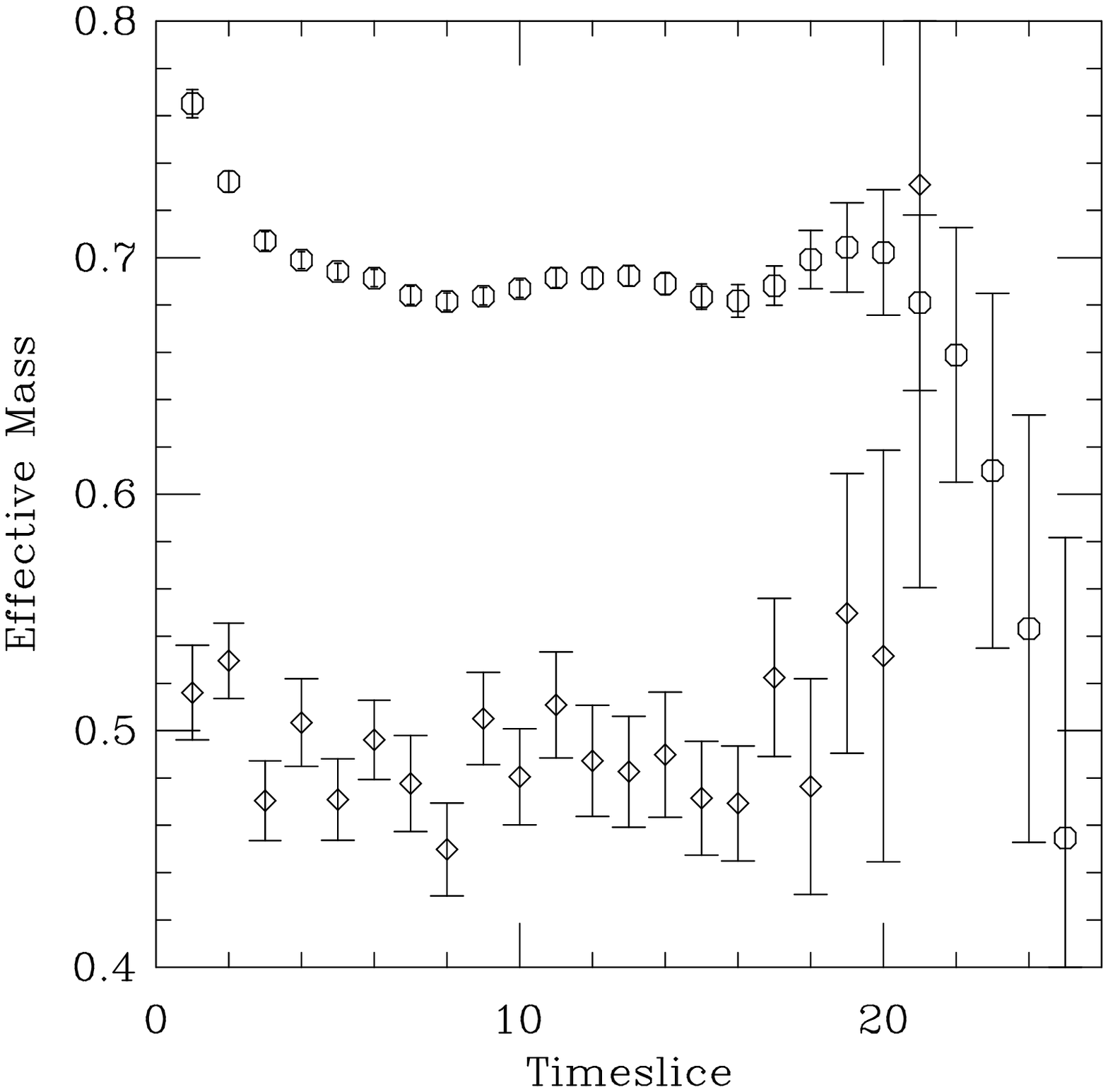}{80mm}
\ewxy{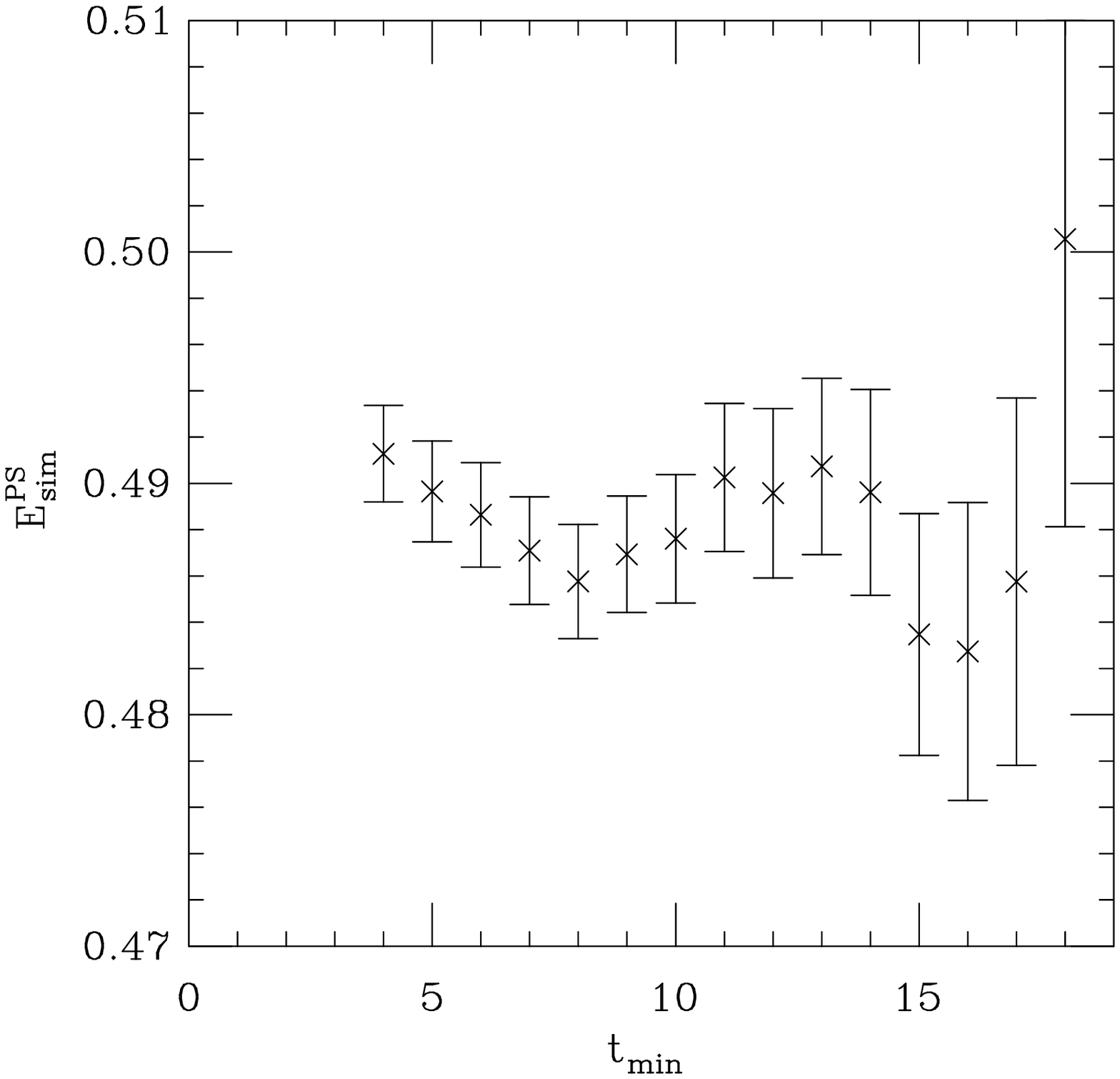}{80mm}}
\vspace{0.5cm}
\centerline{\ewxy{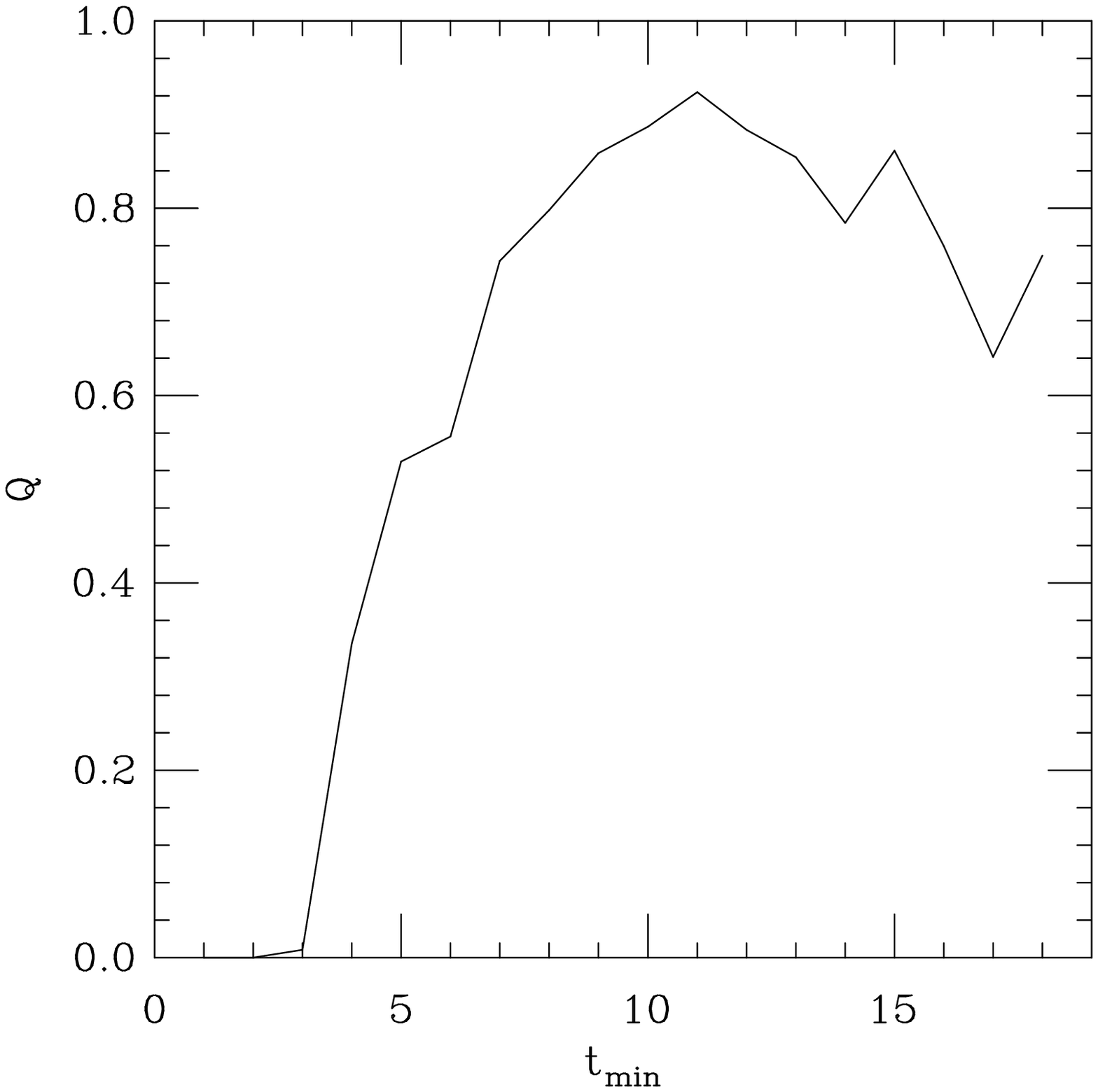}{80mm}}
\caption{The effective masses of the pseudoscalar $C(1,l_{{\cal O}_1})$~(circles) and
$C(1,1)$~(diamonds) correlators, and the corresponding
ground state energy extracted as a function of $t_{min}$ from a
$n_{exp}=1$ vector fit for $aM_0=1.0$ and $\kappa_l=0.1585$. The
effective mass of the $C(1,l_{{\cal O}_1})$ correlator is offset by
$+0.2$. Also shown are the values of $Q$ for the vector fit;
$E_{sim}^{PS}$ is only presented for those values of $t_{min}$
which satisfy $Q>0.1$. $t_{max}$ is fixed at $20$.}
\label{efffit_1}
\end{figure}

To illustrate our results we use the data at $aM_0=1.0$ and
$\kappa_l=0.1585$.  The effective masses of the pseudoscalar
$C(1,l_{{\cal O}_1})$ and $C(1,1)$ correlators, presented
in figure~\ref{efffit_1}, show a clear signal out to approximately 
timeslice $20$. A one exponential~($n_{exp}=1$) simultaneous fit to
both correlators that satisfies the fit criteria
$Q\footnotemark>0.1$\footnotetext{Defined as the one-standard
deviation confidence level.} is possible from an initial
timeslice~($t_{min}$) of $4$ out to timeslice $20$~($t_{max}$). The
variation of $E_{sim}$ with the fit range is $\sim 2\sigma$ for
$t_{min}\gtaeq4$ and this is taken to be the fitting error in our
analysis. A final fitting range of $7-20$ is chosen, for which $Q$ is
judged to be roughly stable to increases in $t_{min}$.

To estimate the remaining excited state contribution to the fitting
parameters extracted the results are compared in table~\ref{cmpfit_1}
with a multi-exponential analysis using a vector and matrix of smeared
correlators~\cite{spec}. The agreement, to within the fitting error,
of the energies and amplitudes provides confidence that we have
isolated the ground state. Since the overlaps of both $C(1,l_{{\cal
O}_1})$ and $C(1,1)$ with the first excited state are
small, $n_{exp}=2$ fits could not be performed successfully to these
correlators.

%
%

\begin{table}[hbpt]
\begin{center}
\begin{tabular}{|c|c|c|c|c|c|c|}\hline
Data & $n_{exp}$ & fit range & Q & $aE_{sim}^{PS}$ & $a^3Z_{l}Z_1$ & $a^{3/2}Z_{1}$\\\hline
1l,11 & 1 & 7-20 & 0.7 & 0.487(2) & 1.09(2) & 10.1(1) \\\hline
1l,2l & 2 & 3-20 & 0.3 & 0.484(3) & 1.02(3) & - \\\hline
11,12,21,22 & 2 & 3-20 & 0.4 & 0.485(2) & - & 10.0(1)\\\hline
\end{tabular}
\caption{The fit parameters, ground state energies and amplitudes extracted 
from vector and matrix fits to the pseudoscalar meson
correlators~(without the current correction) for $aM_0=1.0$ and
$\kappa_l=0.1585$.
\label{cmpfit_1}}
\end{center}
\end{table}

\begin{figure}
\centerline{\ewxy{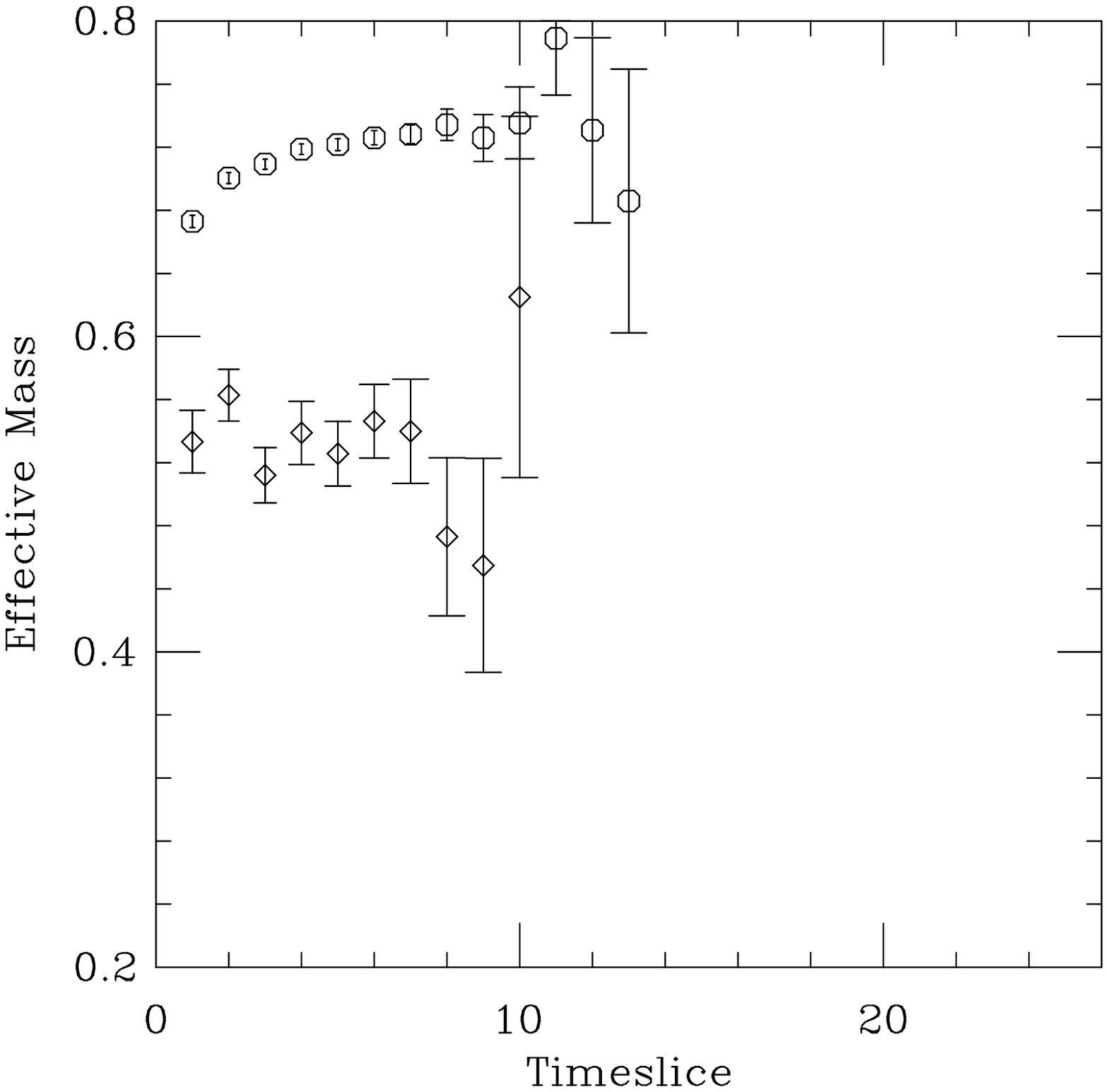}{80mm}
\ewxy{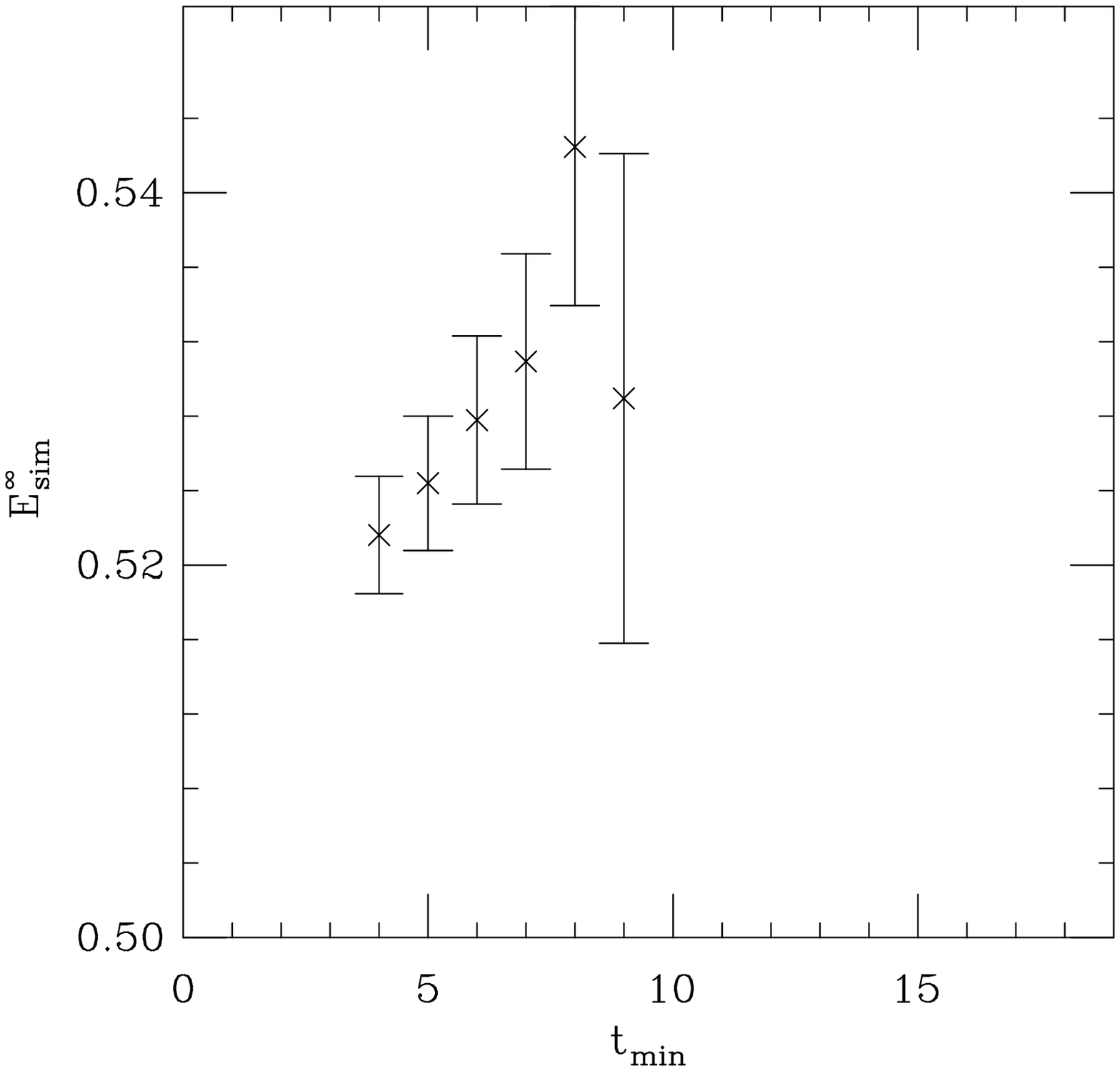}{80mm}}
\vspace{0.5cm}
\centerline{\ewxy{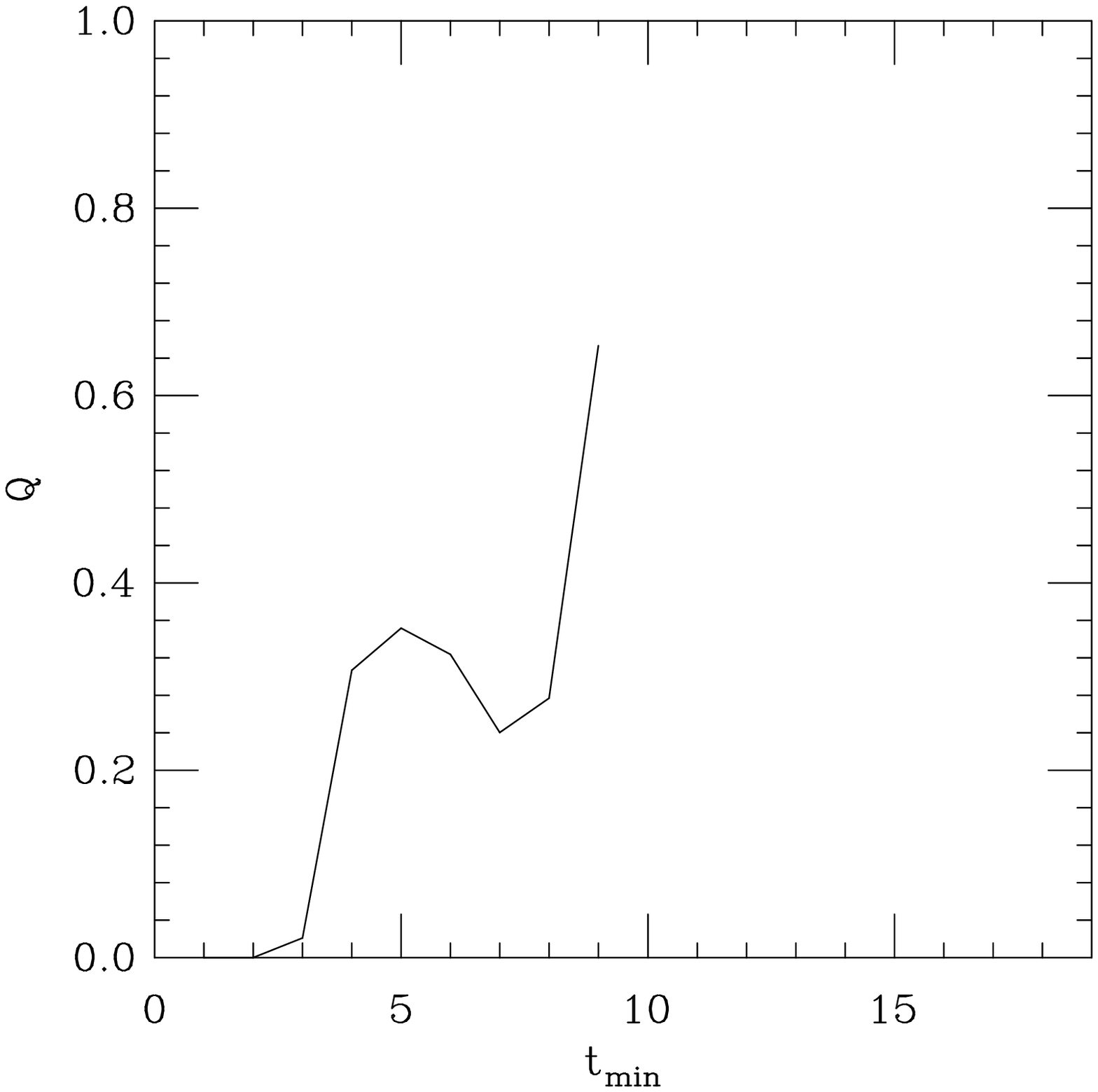}{80mm}}
\caption{The effective masses of the pseudoscalar 
$C(1,l_{{\cal O}_1})$~(circles) and $C(1,1)$~(diamonds)
correlators, and the corresponding ground state energy extracted as a
function of $t_{min}$ from a $n_{exp}=1$ vector fit for $aM_0=\infty$
and $\kappa_l=0.1585$. The effective mass of the $C(1,l_{{\cal O}_1})$
correlator is offset by $+0.2$. Also shown are the values of $Q$ for
the vector fit. $t_{max}$ is fixed at $11$.}
\label{efffit_static}
\end{figure}

\begin{figure}
\centerline{\ewxy{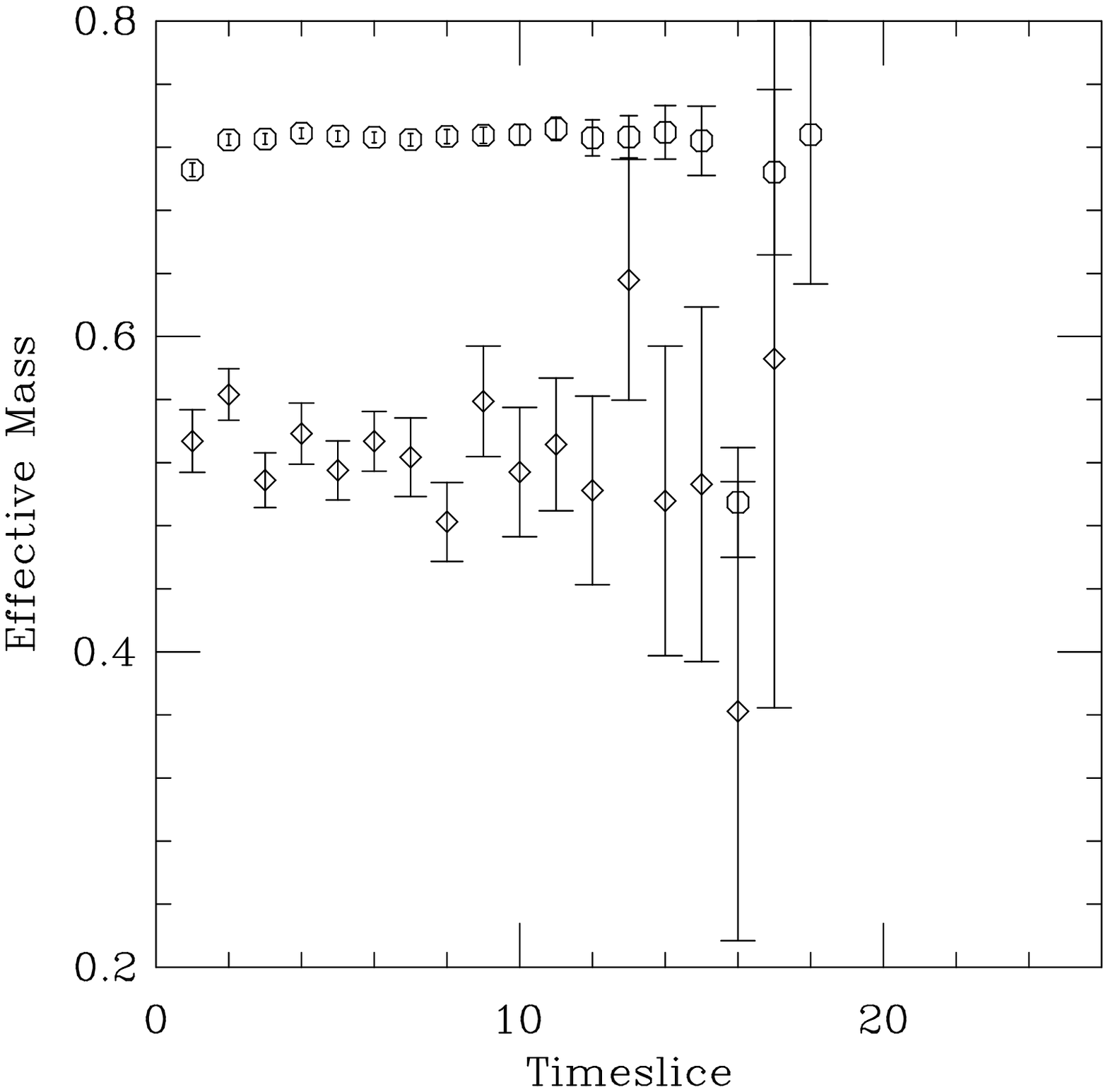}{80mm}}
\caption{The effective masses 
of the pseudoscalar $C(1,l_{{\cal O}_1})$~(circles) and
$C(1,1)$~(diamonds) meson correlators for $aM_0=10.0$ and
$\kappa_l=0.1585$.}
\label{eff_10}
\end{figure}

In the static case, it is more difficult to obtain a satisfactory fit.
From figure~\ref{efffit_static} the signal for the $C(1,1)$ dies away
as $C(1,l_{{\cal O}_1})$ appears to be reaching a plateau.  While a
$n_{exp}=1$ fit to both correlators is possible from $t_{min}=4$ for
$Q>0.1$, figure~\ref{efffit_static} shows the ground state energy is
not stable to changes in the fitting range.  In addition, it was not
possible to bootstrap the fits for $t_{min}<9$.  Thus, we choose
$9-11$ for which $E_{sim}$ is consistent with the results from
different fitting ranges.  Agreement is found when comparing with the
energy and amplitudes obtained from fits to other combinations of
correlators, detailed in table~\ref{cmpfit_static}. However, a better
tuned ground state smearing function would enable the fit parameters
to be extracted with greater confidence. As noted in
reference~\cite{spec}, even for the heaviest finite heavy quark mass
value, $M_0=10.0$ shown in figure~\ref{eff_10}, NRQCD does not suffer
from the same $signal/noise$ problems.

%
%

\begin{table}[hbpt]
\begin{center}
\begin{tabular}{|c|c|c|c|c|c|c|}\hline
Data & $n_{exp}$ & fit range & Q & $aE_{sim}^\infty$ & $a^3Z_1Z_l$ & $a^{3/2}Z_1$\\\hline
1l,11 & 1 & 9-11 & 0.65 & 0.529(13) & 2.07(20) & 10.7(5) \\\hline
1l,2l & 1 & 8-15 & 0.2 & 0.530(9) & 2.1(1) & - \\\hline
11 & 1 & 3-11 & 0.6 & 0.527(6) & - & 10.4(2) \\\hline
\end{tabular}
\caption{The fit parameters, ground state energies and amplitudes extracted 
from vector fits to the pseudoscalar meson correlators for
$aM_0=\infty$ and $\kappa_l=0.1585$.
\label{cmpfit_static}}
\end{center}
\end{table}
%

%
%

The heavy quark is expected to be almost a spectator in the
heavy-light meson and so the wavefunction changes very little with
$M_0$ or the spin content of the meson.  We find it is sufficient to
use the same fitting range, $7{-}20$, for the pseudoscalar mesons for
all finite values of the heavy quark mass.  This fitting range is also
used for the vector mesons, apart from $M_0=0.8{-}1.2$ where it was
necessary to use $9{-}20$. The corresponding values of $E_{sim}$ and
the pseudoscalar and vector decay constants~(without the current
correction) are given in table~\ref{all_m0} for $\kappa_l=0.1585$. The
pseudoscalar meson mass, $M_{PS}$, also shown in the table, is
calculated using the mass shifts, $M_{PS}{-}E_{sim}$, given in
reference~\cite{spec}.  Repeating the analysis for the results at
$\kappa_l=0.1600$, we found $7-20$ is the optimal fitting range for
all $M_0$.

\begin{table}[t]
\begin{center}
\begin{tabular}{|c|c|c|c|c|c|}\hline
$aM_0$ & $E_{sim}$ & $aM_2$ & $a^{3/2}(f\sqrt{M})_{PS}^{uncorr}$ & $a^{3/2}\delta(f\sqrt{M})_{PS}$ & $a^{3/2}(f\sqrt{M})_{V}^{uncorr}$  \\\hline
0.8 & 0.475(3) & 1.36(2) & 0.180(3) & -0.0605(13) & 0.207(5) \\\hline
1.0 &0.488(3) & 1.58(3) & 0.191(3) & -0.0529(10) & 0.179(5) \\\hline
1.2 & 0.497(3) & 1.76(2) & 0.199(3) & -0.0470(9) & 0.188(5) \\\hline
1.7 & 0.510(3) &2.27(2) & 0.215(3) & -0.0373(8) & 0.205(5) \\\hline
2.0 & 0.514(3) &2.58(2) & 0.224(3) & -0.0337(8) & 0.216(5) \\\hline
2.5 & 0.519(4) &3.09(2) & 0.237(4) & -0.0289(6) & 0.228(5) \\\hline
3.0 & 0.522(3) & 3.52(3) & 0.247(5) & -0.0255(8) & 0.239(5) \\\hline
3.5 & 0.524(3) & 4.16(3) & 0.251(5) & -0.0224(6) & 0.244(5) \\\hline
4.0 & 0.525(3) &4.64(4) & 0.258(5) & -0.0204(6) & 0.252(5) \\\hline
7.0 & 0.528(4) & 7.13(12) & 0.285(5) & -0.0132(3) & 0.282(6) \\\hline
10.0 & 0.527(4) & 9.95(12) & 0.300(6) & -0.0099(3) & 0.297(8) \\\hline
$\infty$ & 0.528(5)  &   -      &  0.341(37)  &  -     & - \\\hline
\end{tabular}
\caption{The pseudoscalar ground state energy,  meson mass, 
decay constant and current correction,
and the vector decay constant for all $aM_0$ and $\kappa_l=0.1585$.
Note that the renormalisation
factors are not included.\label{all_m0}}
\end{center}
\end{table}

\begin{figure}
\centerline{\ewxy{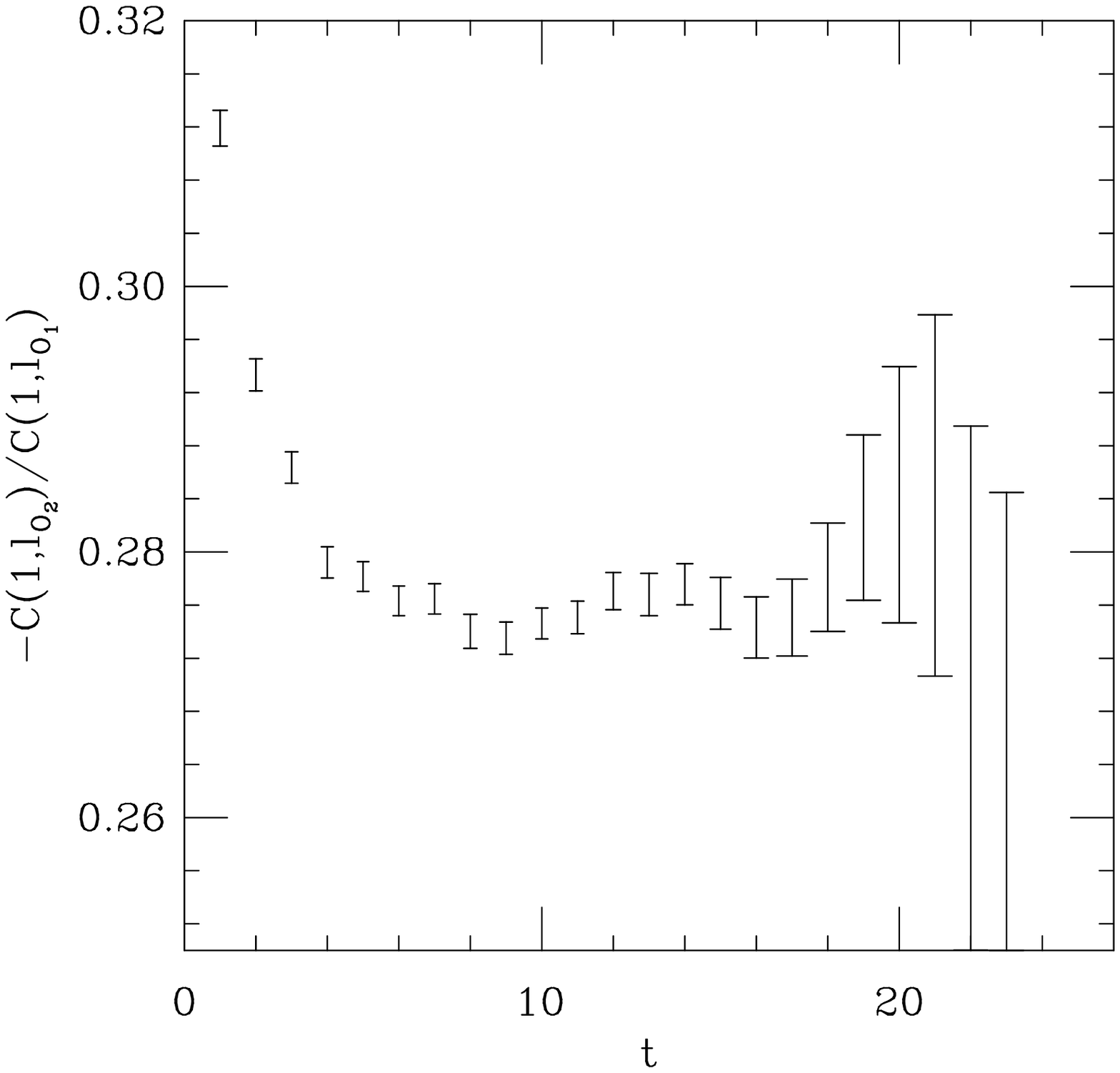}{80mm}
\ewxy{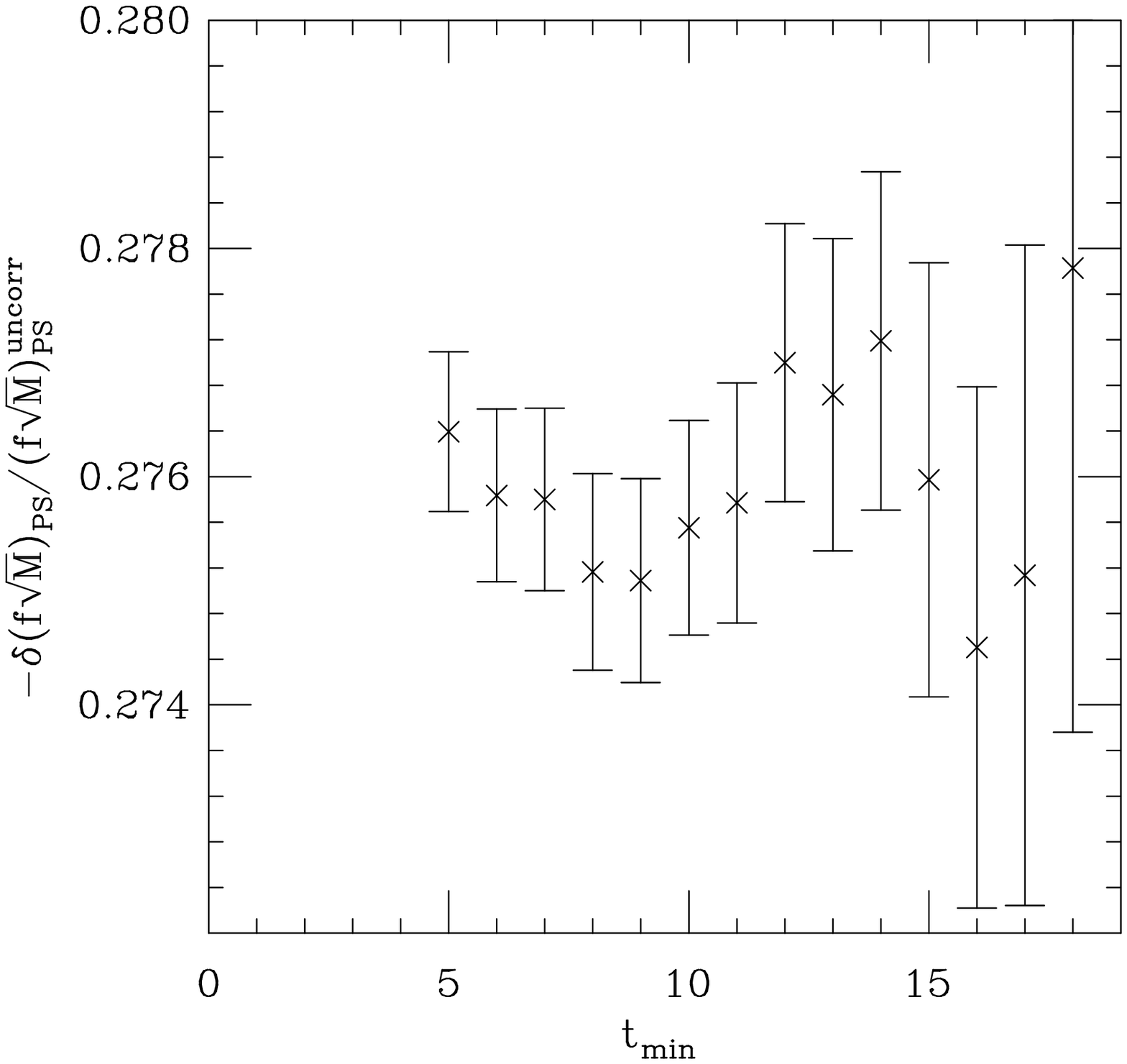}{80mm}}
\vspace{0.5cm}
\centerline{\ewxy{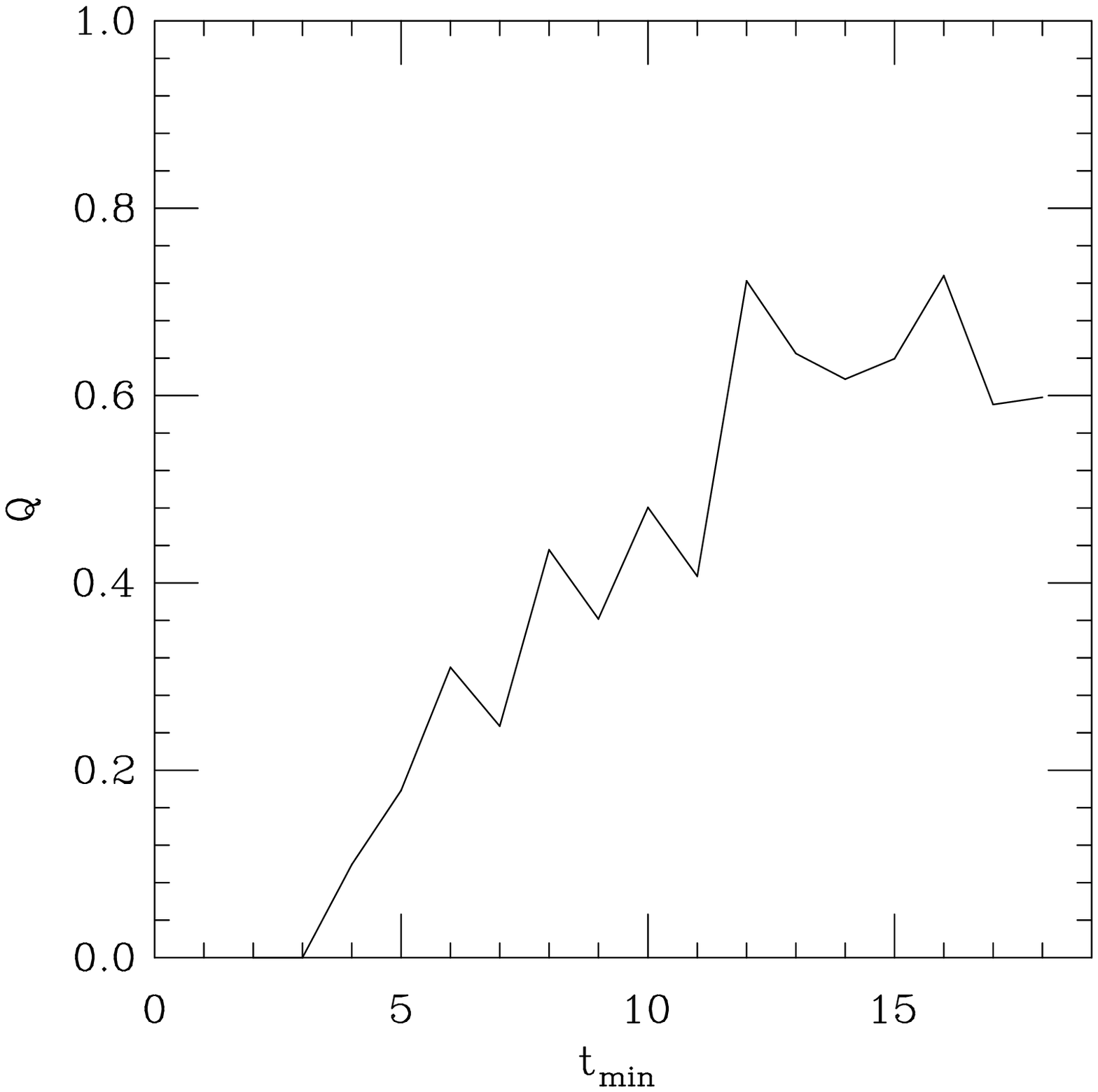}{80mm}}
\caption{The jackknife ratio of the pseudoscalar $C(1,l_{{\cal O}_2})$ and
$C(1,l_{{\cal O}_1})$ correlators and the corresponding amplitude,
$\delta(f\protect\sqrt{M})_{PS}/(f\protect\sqrt{M})_{PS}^{uncorr}$, as
a function of $t_{min}$ obtained from a constant fit for $aM_0=1.0$ and
$\kappa_l=0.1585$. The values of $Q$ for the fits are also shown. $t_{max}$ is
fixed at $20$.}
\label{efffit_corr}
\end{figure}

The current correction is extracted separately by taking the jackknife
ratio of the pseudoscalar $C(1,l_{{\cal O}_2})$ and $C(1,l_{{\cal
O}_1})$ correlators.  In the limit of large times, the ratio tends to a
constant:
\begin{equation}
\frac{C(1,l_{{\cal O}_2})}{C(1,l_{{\cal O}_1})} = \frac{\delta(f\sqrt{M})_{PS}}{(f\sqrt{M})^{uncorr}_{PS}}.
\end{equation}
This ratio is shown in figure~\ref{efffit_corr} for $aM_0=1.0$ and
$\kappa_l=0.1585$.  While a plateau seems to be reached around
timeslice $5$, $Q$ does not stabilise until $t_{min}=12$. Thus, we
choose a fitting range of $12-20$; this was found to be optimal for
all $M_0$. In addition, picking an earlier $t_{min}$ leads to
difficulties when investigating the heavy quark mass dependence of the
current correction.  The statistical uncertainty becomes so small that
terms above $O(1/M^4)$ in the heavy quark expansion of this quantity
are required, and it becomes difficult to extract the coefficients
reliably. Using the previous results for $(f\sqrt{M})_{PS}$ we obtain
the values of $\delta(f\sqrt{M})_{PS}$ shown in table~\ref{all_m0} for
$\kappa_l=0.1585$. It is clear that as $\delta(f\sqrt{M})_{PS}$ is a
$\sim 15\%$ correction to $(f\sqrt{M})_{PS}^{uncorr}$ around
$M_b$~($\sim 2.0$), it is essential to include the current correction
when calculating $f_B$ and investigating its corresponding heavy quark
mass dependence, especially when other sources of error from light 
quarks etc are under control.

%
%

\begin{figure}
\centerline{\ewxy{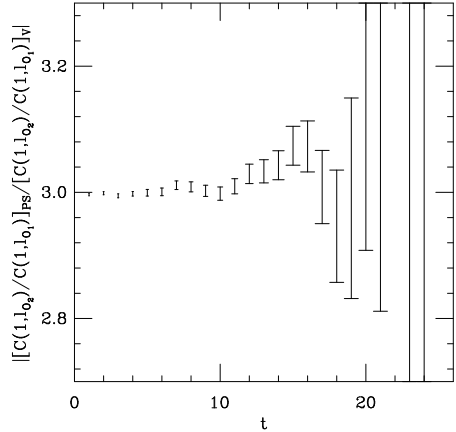}{80mm}}
\caption{The jackknife ratio of $C(1,l_{{\cal O}_2})/C(1,l_{{\cal O}_1})$
for the vector and pseudoscalar mesons for $aM_0=1.0$ and
$\kappa_l=0.1585$.}
\label{rat_ps_v}
\end{figure}

As stated in equations~\ref{fsqrtmq} and~\ref{slope}, group theory
relates the spin dependent contributions to the slopes of the vector
and pseudoscalar decay constants in the continuum. In terms of the
current correction:
\begin{equation}
\left|\frac{ \delta(f\sqrt{M})_{PS}/(f\sqrt{M})_{PS}^{uncorr}} { \delta(f\sqrt{M})_{V}/(f\sqrt{M})_{V}^{uncorr}}\right| = 3.
\label{ratrat}
\end{equation}
On the lattice this relation only holds in the limit of infinite
statistics. Figure~\ref{rat_ps_v} shows $\left|\frac{ [C(1,l_{{\cal
O}_2})/C(1,l_{{\cal O}_1})]_{PS}} { [C(1,l_{{\cal O}_2})/C(1,l_{{\cal
O}_1})]_{V}}\right|$ for $aM_0=1.0$ and $\kappa_l=0.1585$. In the limit of
large times this ratio of correlators tends to the R.H.S. of
equation~\ref{ratrat}. We find equation~\ref{ratrat} is satisfied to
within $\sim 1{-}2 \sigma$ and thus, we can further reduce the number of
matrix elements that need to be calculated by using $-\frac{1}{3} [
\delta (f\sqrt{M})_{PS}/(f\sqrt{M})_{PS}^{uncorr}]
(f\sqrt{M})_V^{uncorr}$ for the vector current correction.

We compare $E_{sim}^{PS}$ with \mbox{$2M_0\bra{0} {\cal
O}_2^{PS}\ket{PS}_{NRQCD}/ \bra{0} {\cal O}_1^{PS}
\ket{PS}_{NRQCD}$} in table~\ref{cmp_esim_corr}. The 
$20\%\sim O(\alpha_S)$ disagreement between these two quantities is
approximately independent of heavy quark mass, and hence is most
likely to be due to the omission of the lattice renormalisation
factors rather than $O(1/M)$ terms. Note that both quantities are
roughly a factor of $2$ larger than $\bar{\Lambda}$ i.e. quite
different from the analogous quantities in HQET.

The spin-average of the vector and pseudoscalar decay constants, as
well as the ratio, with and without the current correction were
computed.  The results are given in table~\ref{manip_allm0}. In order
to investigate the heavy quark mass dependence of these quantities,
detailed in the next section, 100 bootstrap ensembles were generated
for each correlated fit. The bootstrap averages of the fit parameters
are consistent with the corresponding unbooted values given in the
tables. In the static case we obtain,
\begin{equation}
aE_{sim}^\infty = 0.559(37)\hspace{0.5cm} a^3Z_lZ_1 = 2.82(88)
\hspace{0.5cm} a^{3/2} Z_1 = 12.4(1.7).
\end{equation}
These values are consistent, but with much larger errors than those
obtained from unbooted data.  This underscores the $signal/noise$
problems and the rough nature of our determinations of
$(f\sqrt{M})^{static}$. The corresponding estimate of
$(f\sqrt{M})^{stat} = 0.4(1)$ is so uncertain as not to be useful.
Since the purpose of our static result is to compare with the more
accurate extrapolation to the infinite mass limit of the NRQCD
results, a rough value is sufficient and we use a value consistent
with the unbooted fits, $(f\sqrt{M})^{stat} = 0.341(37)$, which has
smaller error bars.

%
%
\begin{table}[hbpt]
\begin{center}
\begin{tabular}{|c|c|c|}\hline
$aM_0$ & $aE_{sim}$ & $-2M_0a\frac{\delta(f\sqrt{M})_{PS}}{(f\sqrt{M})_{PS}^{uncorr}}$\\\hline
0.8 & 0.475(3)  &  0.537(2) \\\hline
1.0 & 0.488(3)  &  0.554(3) \\\hline
1.2 & 0.497(3)  &  0.567(3) \\\hline
1.7 & 0.510(3)  &  0.589(3) \\\hline
2.0 & 0.514(3)  &  0.600(3) \\\hline
2.5 & 0.519(4)  &  0.612(4) \\\hline
3.0 & 0.522(3)  &  0.621(4) \\\hline
3.5 & 0.524(3)  &  0.627(4) \\\hline
4.0 & 0.525(3)  &  0.631(5) \\\hline
7.0 & 0.528(4)  &  0.649(6) \\\hline
10.0 &0.527(4)  &  0.657(8) \\\hline
\end{tabular}
\caption{A comparison of the pseudoscalar ground S-state energy and
$\bra{0} {\cal O}_2 \ket{PS}_{NRQCD}/\bra{0} {\cal O}_1
\ket{PS}_{NRQCD}$ for all $aM_0$ and $\kappa_l=0.1585$. Note that the
renormalisation factors are not included.
\label{cmp_esim_corr}}
\end{center}
\end{table}
\begin{table}[t]
\begin{center}
\begin{tabular}{|c|c|c|c|}\hline
$aM_0$ &  $a^{3/2}(\overline{f\sqrt{M}})$ & $(\frac{(f\sqrt{M})_{PS}}{(f\sqrt{M})_V})^{uncorr}$ & $(\frac{(f\sqrt{M})_{PS}}{(f\sqrt{M})_V})^{tot}$ \\\hline
0.8 &  0.172(5) & 1.070(18) & 0.641(11)\\\hline
1.0 &  0.182(5) & 1.065(19) & 0.706(13)\\\hline
1.2 &  0.190(5) & 1.059(18) & 0.751(13)\\\hline
1.7 &  0.208(3) & 1.047(7) & 0.819(6)\\\hline
2.0 &  0.217(5) & 1.042(7) & 0.845(6)\\\hline
2.5 &  0.229(5) & 1.037(7) & 0.875(5)\\\hline
3.0 &  0.241(5) & 1.032(7) & 0.895(5)\\\hline
3.5 &  0.246(5) & 1.028(6) & 0.910(5)\\\hline
4.0 &  0.253(5) & 1.025(7) & 0.920(6)\\\hline
7.0 &  0.282(6) & 1.013(5) & 0.952(4)\\\hline
10.0 & 0.298(8) & 1.007(4) & 0.964(5)\\\hline
\end{tabular}
\caption{The spin-averaged decay constant, and the ratio of the
pseudoscalar to vector decay constant, both with and without the
current correction for all $aM_0$ and $\kappa_l=0.1585$. Note that the renormalisation
factors are not included.
\label{manip_allm0}}
\end{center}
\end{table}

\begin{figure}
\centerline{\ewxy{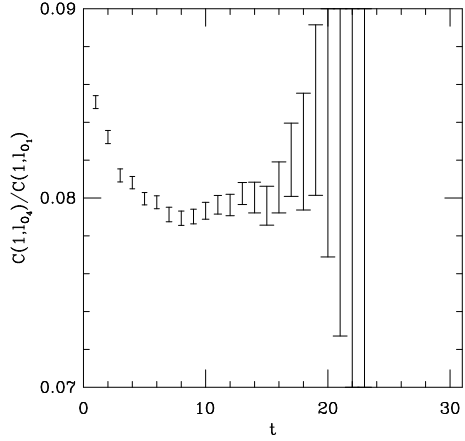}{80mm}}
\caption{The jackknife ratio of $C(1,l_{{\cal O}_4})/C(1,l_{{\cal O}_1})$
for the vector meson at $aM_0=1.0$ and $\kappa_l=0.1585$.}
\label{corr_vec}
\end{figure}

While the one loop correction to the decay constant is not included in
this analysis, the size of the vector matrix element $\bra{0} {\cal
O}_4
\ket{V}_{NRQCD}$ is shown in figure~\ref{corr_vec} as a percentage of
$\bra{0} {\cal O}_1 \ket{V}_{NRQCD}$ for $aM_0=1.0$ and
$\kappa_l=0.1585$. This is roughly the same size as the tree level
matrix element,
\begin{equation}
\delta(f\sqrt{M})_V/(f\sqrt{M})_V^{uncorr}=
-\frac{1}{3}\delta(f\sqrt{M})_{PS}/(f\sqrt{M})_{PS}^{uncorr}\sim
0.28/3.
\end{equation}
Since the one loop contributions should be suppressed by a
factor of $O(\alpha_S) \sim 20\%$, the contribution to the vector
decay constant is around $2\%$ of \\\mbox{$\bra{0} {\cal O}_1
\ket{V}_{NRQCD}$}. This is of order $\sim 2\sigma$ in the statistical
errors of $(f\sqrt{M})_{V}^{uncorr}$ at this mass. The one loop
correction to the pseudoscalar decay constant is similarly small.

\subsection{Heavy quark mass dependence of the decay constants}
\label{hq_decay}
\begin{figure}
\centerline{\ewxy{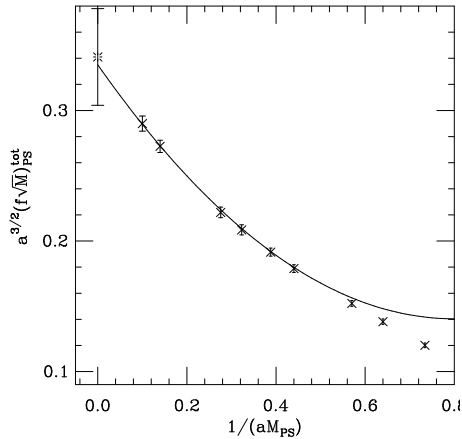}{80mm}}
\caption{$a^{3/2}(f\protect\sqrt{M})_{PS}^{tot}$ vs $1/(aM_{PS})$ 
for $\kappa_l=0.1585$. The solid line indicates a quadratic fit to the
heaviest six points and $1/(aM_B)\sim 0.4$. The static data point,
shown as a burst, is not included in the fit. Note that the renormalisation
factors are not included.}
\label{decay_ps_corr}
\end{figure}

The decay constants have been computed over a wide range of heavy
quark mass and this is ideal for investigating heavy quark symmetry.
Figure~\ref{decay_ps_corr} presents the results for
$(f\sqrt{M})_{PS}^{tot}$ as a function of $1/M_{PS}$ for
$\kappa_l=0.1585$. The static result is also included in the plot. A
correlated fit to the data was performed using the functional form:
\begin{eqnarray}
f\sqrt{M} & = & C_0 + \frac{C_1}{M} + \frac{C_2}{M^2} + \frac{C_3}{M^3}.
\end{eqnarray}
The fitting range was varied keeping the starting point fixed at the
heaviest data point, $aM_0=10.0$.  Beginning with a linear function,
the fitting procedure was repeated adding quadratic and cubic terms.
In general, in order to have confidence in the value for $C_1$ a
quadratic fit is required. Note that in reference~\cite{spec} we found
when performing correlated fits to the results as a function of $1/M$
the values of $n$ used in the evolution equation for $aM_0=3.5$ and
$4.0$ are too low; the results for these data points are not included
in any of the fits.

Table~\ref{fit_ps} summarises the coefficients extracted from various
fits. A `good fit' is defined in the same way as for the propagator
fits: $Q>0.1$. The fitting range for which $Q$ falls below this
cut-off is taken to signify when higher terms can be resolved. Thus,
we do not present fits including an additional term over a smaller
fitting range.  The decay constant is only consistent with a linear
dependence on $1/M_{PS}$ for the first three data points,
for which $1/M_{PS}\ltaeq 0.3$. The quadratic term is needed for
$1/M_{PS}\ltaeq 0.6$, i.e. there are significant $O(1/M^2)$
contributions to the decay constant in the region of the $B$
meson~($1/(aM_{PS})\sim0.4$). Since only the $O(1/M)$
contributions are correctly included in this analysis it is necessary
to go to the next order in the NRQCD expansion to calculate $f_B$
accurately. A fit to all the data points is possible including the
cubic term.

\begin{table}[hbpt]
\begin{center}
\begin{tabular}{|c|c|c|c|c|c|c|c|}\hline
Order & fit range &n.d.o.f. & Q & $a^{3/2}C_0$ & $a^{5/2}C_1$ & $a^{7/2}C_2$ & $a^{9/2}C_3$\\\hline
1 & 1-3 &1 & 0.3 & 0.324(6) & -0.371(22) & - & - \\\cline{2-8}
  & 1-4 &2 & 0.0 & 0.317(6) & -0.329(15) & - & - \\\hline
2 & 1-4 &1 &0.7 & 0.342(10) & -0.561(71) & 0.458(137) & - \\\cline{2-8}
  & 1-5 &2 &0.7 & 0.338(10) & -0.525(55) & 0.379(75) & - \\\cline{2-8}
  & 1-6 &3 &0.7 & 0.335(9) & -0.486(40) & 0.304(60) & - \\\cline{2-8}
  & 1-7 &4 &0.7 & 0.332(7) & -0.456(23) & 0.254(26) & - \\\hline
 3 & 1-8 &4& 0.8 &0.346(9) & -0.623(52) & 0.740(119) &-0.425(104) \\\cline{2-8}
  & 1-9 &5 & 0.9 &0.346(8) & -0.614(45) & 0.716(100) &-0.406(69) \\\hline
\end{tabular}
\caption{The coefficients extracted from fits to 
$a^{3/2}(f\protect\sqrt{M})_{PS}^{tot}$ as a function of $1/(aM_{PS})$ for
$aM_0=1.0$ and $\kappa_l=0.1585$. Note that the renormalisation
factors are not included. The fit range $1{-}3$ denotes a fit to
the data at $aM_0=10.0$, $7.0$ and $3.0$. The data at $aM_0=4.0$ and $3.5$
are not included in the fit.\label{fit_ps}}
\end{center}
\end{table}

A quadratic fit to the heaviest six data points is shown in
figure~\ref{decay_ps_corr}. The extrapolation of the NRQCD results at
finite heavy quark mass is clearly consistent with the static data
point. Since the static theory is the heavy quark mass limit of NRQCD
this is expected and we assume this is not spoilt by the inclusion of
the renormalisation factors.  A striking feature of the plot is the
very large slope and the corresponding large deviations from the
static limit of $\sim 50\%$ around $1/M_B$.

Note that with only the last two data points requiring a cubic fit,
the coefficient $C_1$ is not as well determined as for the quadratic
fit. Hence, we only consider the quadratic fits and find
$a^{3/2}C_0=0.34(1)$. Similarly, $a^{5/2}C_1\sim -0.46(5)$.  Thus, we
arrive at $ac_{P}=aC_1/C_0\sim -1.35(15)$, much larger than the naive
expectation of $O(a\Lambda_{QCD})\sim -0.2$. Converting to physical
units, $c_{P}\sim -2.8(5)$~GeV, using $a^{-1}=1.8{-}2.4$. We found the
slope parameters do not depend significantly on the light quark mass.

This slope is significantly larger than that found in previous lattice
simulations which found $c_{P}\sim -0.8$~GeV~\cite{ukqcd} and
$-1.14$~GeV~\cite{milc}. We believe the discrepancy is due to previous
results being obtained by extrapolation from around the $D$ meson with
too small a range of masses when using Wilson or Clover heavy quarks;
consistency between the static limit and results in this region
has not been shown. In addition, our results indicate the
deviations from the static limit may be so large at $M_D$ that a heavy
quark expansion is not valid in this region.

Onogi et al, using the Fermilab approach to an order equivalent to
$O(1/M)$ in NRQCD, also find a slope around $-1$ GeV~\cite{fermilab};
data points around the $B$ and $D$ meson are used.  The pseudoscalar
renormalisation factors for this method have not yet been calculated
and a proper comparison of the extrapolation of the results with the
static limit cannot yet be made. Consistency should be found between this
method and NRQCD, and it is important to confirm the slope in the
large mass region, $M_{PS} \gg M_B$.

\begin{figure}
\centerline{\ewxy{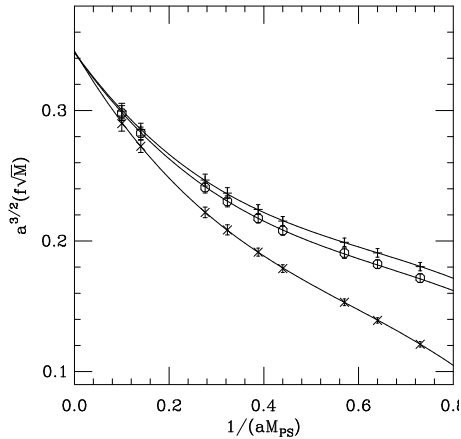}{80mm}}
\caption{The pseudoscalar decay constant both with~(crosses) and
without~(pluses) the current correction and the spin-average decay
constant~(circles) as a function of $1/(aM_{PS})$ for
$\kappa_l=0.1585$. The solid lines indicate cubic fits to all the data
points. Note that the renormalisation
factors are not included.}
\label{decay_all}
\end{figure}

To examine the large slope of the decay constant further we obtained
the individual contributions to the pseudoscalar and vector matrix
elements from each of the $O(1/M)$ interactions.
Figure~\ref{decay_all} shows $(\overline{f\sqrt{M}})$,
$(f\sqrt{M})_{PS}^{tot}$ and $(f\sqrt{M})_{PS}^{uncorr}$ as a
function of $1/M_{PS}$. From section~\ref{hqsym} the slope of the
spin-average decay constant depends only on the kinetic energy of the
heavy quark, while the other two quantities include contributions from
the spin dependent interactions. It is clear from the large mass
region where the slope is linear that $G_{kin}$ is much greater than
$G_{hyp}$ or $G_{corr}$, and is the source of the large slope.
Qualitatively, the spin dependent terms are expected to be suppressed
since they break spin as well as flavour symmetry. Performing
correlated fits to $(\overline{f\sqrt{M}})$, detailed in
table~\ref{coeffs}, we find $aG_{kin}=C_1/C_0\sim-1.26(15)$ or
$-2.6(5)$~GeV.

This result agrees with Hashimoto~\cite{hash}, who calculates the
pseudoscalar decay constant at $\beta=6.0$ in the quenched
approximation and finds the coefficient of the slope at $O(1/M)$
roughly $-2.4(1.1)$~GeV.  Since the correction to the current is
omitted in his calculation, this coefficient corresponds to $G_{kin} +
6G_{hyp}$.

Note that even though only $O(1/M)$ terms are included in the NRQCD
action we see quadratic and cubic behaviour in the decay constant in
figure~\ref{decay_ps_corr}; this is due to higher powers of the
existing $O(1/M)$ terms. These higher order powers are dominated by
powers of the kinetic energy terms as is clear from
figure~\ref{decay_all}, which shows that the spin-dependent and
spin-independent slopes of $f\sqrt{M}$ show little
difference even at relatively small values of $M$.  Also, the missing
$O(1/M^2)$ terms in the NRQCD action are all spin-dependent so they
cannot contribute to the spin-independent slope at $O(1/M^2)$. We
would therefore expect that our spin-independent slope was correct
even through $O(1/M^2)$ except that there are spin-independent matrix
element corrections at this order. We do not know the size of these
corrections, but it is unlikely that they would be larger than the
spin-independent $1/M^2$ terms that are present in our calculation
already. Later we will allow for this by taking a rather pessimistic
$100\%$ error in our $1/M^2$ coefficient.

\begin{table}[hbpt]
\begin{center}
\begin{tabular}{|c|c|c|}\hline
 &  $a^{3/2}C_0$ & $a^{5/2}C_1$ \\\hline
$a^{3/2}(\overline{f\sqrt{M}})$ &  0.335(7) & -0.43(5) \\\hline
$(\frac{(f\sqrt{M})_{PS}}{(f\sqrt{M})_V})^{uncorr}$ &  0.996(5) & 0.11(1) \\\hline
$(\frac{(f\sqrt{M})_{PS}}{(f\sqrt{M})_V})^{tot}$ &  1.009(5) & -0.40(4) \\\hline
 & $aC_0$  & $a^2C_1$\\\hline
$2M_0a\frac{\delta(f\sqrt{M})_{PS}}{(f\sqrt{M})_{PS}^{uncorr}}$ & -0.675(7) &0.19(1)  \\\hline
\end{tabular}
\caption{The coefficients extracted from fits to combinations of the
pseudoscalar and vector decay constants as a function of $1/(aM_{PS})$
for $\kappa_l=0.1585$. The errors include the variation in the coefficients
using different orders in the fit function.
\label{coeffs}}
\end{center}
\end{table}

\begin{figure}
\centerline{\ewxy{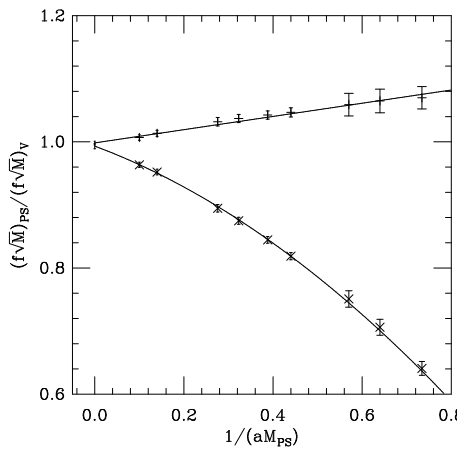}{80mm}}
\caption{The ratio of the pseudoscalar to vector decay constant
both with~(crosses) and without~(pluses) the current correction as a
function of $1/(aM_{PS})$ for $\kappa_l=0.1585$. The solid lines
indicate a linear fit to
$((f\protect\sqrt{M})_{PS}/(f\protect\sqrt{M})_V)^{uncorr}$ and a
quadratic fit to
$((f\protect\sqrt{M})_{PS}/(f\protect\sqrt{M})_V)^{tot}$, where all
data points are included in the fits. The error in the extrapolation
to the static limit is indicated for both fits. Note that the renormalisation
factors are not included.}
\label{decay_rat}
\end{figure}

To isolate the effects of $G_{hyp}$ and $G_{corr}$, consider the ratio of the
pseudoscalar and vector decay constants, with and without the current
corrections, shown in figure~\ref{decay_rat}. Both data sets are
consistent with $1$ in the static limit, as expected. Note that
without the current corrections the ratio, which is determined solely
by the hyperfine interaction, is greater than $1$ indicating $G_{hyp}$
is positive. We find $ac_P''\sim+0.11(1)$ or $.23(4)$~GeV in physical
units, and hence $aG_{hyp} = c_P''/8\sim0.014(1)$ or
$0.029(5)$~GeV. Including $\delta(f\sqrt{M})$, the slope is a physical
quantity and $ac_P'''=-0.40(4)$ or $-0.85(15)$~GeV.

\begin{figure}
\centerline{\ewxy{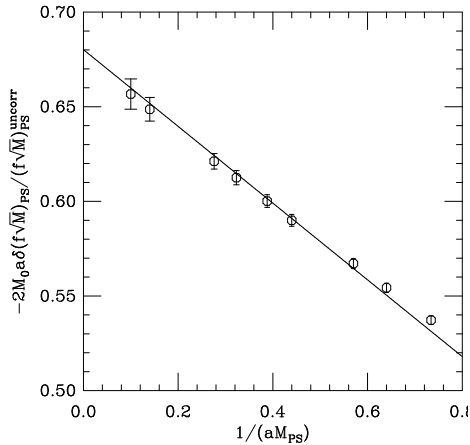}{80mm}}
\caption{$2M_0a\delta(f\protect\sqrt{M})_{PS}/(f\protect\sqrt{M})_{PS}^{uncorr}$ 
vs $1/(aM_{PS})$ for $\kappa_l=0.1585$. The solid line indicates a
linear fit to the six heaviest data points. Note that the renormalisation
factors are not included.}
\label{currfit}
\end{figure}

By subtracting $c_P''$ and $c_P'''$ we estimate
$aG_{corr}\sim-0.77(8)$.  However, the correction to the current can
be extracted directly and more accurately as the intercept of
\mbox{$2M_0\delta(f\sqrt{M})_{PS}/(f\sqrt{M})_{PS}^{uncorr}$}
shown in figure~\ref{currfit}. From table~\ref{coeffs}, we obtain
consistent results, $aG_{corr}=-0.675(7)$ or $-1.42(20)$~GeV. As in
the finite heavy quark mass case, $G_{corr}$ agrees with
$-E_{sim}^\infty=-0.528(5)$ to within approximately $O(\alpha_S)$.

%
%

A comparison can be made with the predictions of QCD sum rules.  The
results obtained by Neubert~\cite{neubert} using HQET, shown in
table~\ref{compneub}, are in good agreement with our analysis. As
noted previously, the unphysical parameters $G_{hyp}$ and $G_{corr}$
differ; from table~\ref{compneub} the former differs in sign as well
as in magnitude with our result.  It remains unclear, however, why the
slope, and thus $G_{kin}$, is much larger than naively expected; a
similar analysis of the spectrum~\cite{spec} showed deviations from
the static limit in agreement with naive expectations.  The decay
constant may be unique in this respect, however, it is clear this
should not be assumed, particularly for quantities such as
semi-leptonic form factors of $B$ mesons which are not protected by
Luke's theorem.

\begin{table}
\begin{center}
\begin{tabular}{|c|c|c|}\hline
 & Sum Rules & NRQCD\\\hline
$c_P$ & -2.9(5)  &  -2.8(5)\\\hline
$c_P'''$ & -0.9(1) & -0.85(15)\\\hline
$G_{kin}$ & -2.3(4)  & -2.6(5)\\\hline
$G_{hyp} $& -0.07(3) & 0.029(5)\\\hline
$G_{corr}$& -0.50(7) & -1.42(20)\\\hline
\end{tabular}
\caption{A comparison in GeV of the decay constant 
slope parameters and coefficients extracted in this simulation with those
computed by Neubert~\protect\cite{neubert} using HQET and QCD sum rules. The
errors are dominated by the uncertainty in $a^{-1}$.
\label{compneub}}
\end{center}
\end{table}

By chirally extrapolating the results and interpolating to $1/(aM_B)$,
we find $f_B\sim126-166$~MeV. The static renormalisation factor,
$Z_A^{stat}=0.73$~\cite{eichten}~\footnote{This renormalisation
constant corresponds to $\tilde{Z}Z_{cont}$ in the notation used in
this reference. Also, the anomalous dimension factor is set to $1$. },
is used, where we guess $aq^*_{Z_A}=2.0$ at $aM_0=\infty$. To estimate
the error in $f_B$ due to the truncation of the NRQCD series we assume
a $100\%$ error in the quadratic coefficients of the heavy quark mass
expansion.  Thus, we find a $\sim 20\%$ uncertainty in $f_B$, which is
comparable to the uncertainties arising from the light quark sector.
This is likely to be an overestimate; since the contributions from the
kinetic energy term in the NRQCD action dominate the slope, the
corrections to the matrix element at $O(1/M^2)$ are unlikely to be
larger than the quadratic contributions already present. Similarly,
from the size of $G_{kin}$ the next order spin-independent terms in
the action, $-p^4/8M^3$, are expected to lead to larger contributions
to $f_B$ than naively predicted, but this is unlikely to be larger
than the lower order contributions. The $O(\alpha_S)\sim 20\%$ error
in the slope introduced by using the static renormalisation factor is
less significant. In the static limit, we obtain
$f_{PS}^{stat}=237{-}295$~MeV at $\kappa_l=0.1585$ which is close to
$\kappa_s$.

Our determination of $f_B$ is consistent with the current world
average of $200(40)$~MeV~\cite{chrisa}. Our analysis is being
performed in conjunction with a simulation using $\beta=6.0$ quenched
configurations~\cite{arifadecay}. With such large systematic errors it
is not possible to discern any difference between the partial
inclusion of dynamical quarks in this analysis and quenched results.
However, the decay constant is related to the wavefunction at the
origin, $f_B\sqrt{M_B}=\sqrt{12}|\psi(0)|$, and thus it is expected to
be sensitive to short distance physics and the quenched approximation.
Booth~\cite{booth} has estimated the effect of quenching to be
$10-15\%$ in $f_B$ and $\sim 5\%$ in $f_{B_s}/f_B$ using chiral
perturbation theory coupled with HQET. However, the prediction is
fairly dependent on the light quark mass.  In the future we aim to
reduce the systematic errors in both our quenched and dynamical
simulations by using tadpole improved clover light fermions and
including higher orders terms in the NRQCD expansion. Thus, we
expect to isolate and remove the full effects of quenching.

With an experimental measurement of mixing also in the
$\bar{B^0_s}{-}B^0_s$ system only the ratio $f_{B_s}B_{B_s}/f_BB_B$ is
required to extract $|V_{ts}/V_{td}|$. Some cancellation of systematic
errors is expected for this ratio, and hence it should be determined
more accurately. Table~\ref{decay_rat_c_s} presents the results for
$((f\sqrt{M})_s/(f\sqrt{M})_d)^{tot}$, where $_s$ and $_d$ denote
results for a light quark mass extrapolated to the strange quark mass
and the chiral limit respectively. The expected $O(m_s/M)$ dependence
on the heavy quark mass is not resolved in our results and the ratio
is independent of $M_0$. The error indicated is purely statistical.
There is an additional $\sim 1 \sigma$ error due to the uncertainty in
$\kappa_s$.

%
%
\begin{table}[hbpt]
\begin{center}
\begin{tabular}{|c|c|c|c|c|c|c|}\hline
$aM_0$ & 0.8& 1.0 &1.2& 1.7& 2.0 & 2.5 \\\hline
$(\frac{(f\sqrt{M})_{s}}{(f\sqrt{M})_{d}})^{tot}$ & 1.26(3) & 1.26(3) & 1.25(3) & 1.25(3) & 1.25(3) & 1.26(4) \\\hline
$aM_0$ & 3.0 & 3.5 & 4.0 & 7.0 & 10.0 & $\infty$ \\\hline
$(\frac{(f\sqrt{M})_{s}}{(f\sqrt{M})_{d}})^{tot}$ &1.26(4) & 1.26(4) & 1.25(4) & 1.24(5) & 1.25(3) & - \\\hline
\end{tabular}
\caption{The ratio of the pseudoscalar decay constant for 
$\kappa_l=\kappa_s$ and $\kappa_l=\kappa_c$ for all $aM_0$, where
$\kappa_s=0.1577$ is obtained using the ratio $M_\phi/M_\rho$. The
errors shown are purely statistical.\label{decay_rat_c_s}}
\end{center}
\end{table}

\section{Conclusions}
\label{conc}
We presented a lattice study, partially including the effects of
dynamical quarks, of the heavy-light pseudoscalar and vector decay
constants using Wilson light quarks and NRQCD heavy quarks. All
$O(1/M)$ terms in the NRQCD action and matrix elements were included;
an $O(a\Lambda_{QCD}/M)$ wave-function renormalization induced by the
evolution equation is the same order as the light quark discretization
errors.  We find consistency between the extrapolation of the NRQCD
result and the static case, as expected, although the static result is
not very well determined.  There are significant $O(1/M^2)$
corrections to $f\sqrt{M}$ around $M_B$, and this leads to a large
systematic error in extracting $f_B$; the use of Wilson light quarks
introduces a systematic error of roughly the same magnitude. It is
necessary to go to the next order in NRQCD and use an improved light
quark action in order to reliably extract $f_B$.

We found the slope of the decay constants with $1/M$ to be
significantly larger than that found in previous lattice
simulations. We believe this is because we are closer to the static
limit. In particular, for the first time we obtained the three
$O(1/M)$ contributions to the linear component of the slope
separately. We found that $G_{kin}$ dominates and thus is responsible
for the large slope. Good agreement is found between our results for
$G_{kin}$ and physical combinations of $G_{hyp}$ and $G_{corr}$ and a
prediction of QCD sum rules. The results show that the kinetic energy
of the heavy quark gives rise to contributions to $f\sqrt{M}$ much
greater than $O(\Lambda_{QCD})$, contrary to naive expectations.
\section{Acknowledgements}
The computations were performed on the CM-2 at SCRI.  We thank the
HEMCGC collaboration for use of their configurations and light quark
propagators. We would like to thank T.~Bhattacharya, R.~Gupta and
A.~Kronfeld for useful discussions. This work was supported by SHEFC,
PPARC and the U.S.~DOE. We acknowledge support by the NATO under grant
CRG~941259 and the EU under contract CHRX-CT92-0051.


\begin{thebibliography}{99}
\bibitem{isgur} J.~M.~Flynn and N.~Isgur, J.~Phys. {\bf G18} 1627 (1992).
\bibitem{neublong} M.~Neubert, Phys. Reports {\bf 245} 259 (1994).
\bibitem{chrisa} C.~Allton, plenary talk presented at the 
{\it International Symposium on 
Lattice Field Theory, Melbourne. Australia, 11-15 July 1995}, to
appear in Nucl.~Phys. {\bf B} (Proc.~Suppl.).
\bibitem{qcdsum} S.~Narison, talk presented at the {\it Cracow Epiphany Conference on Heavy Quarks}, Montpellier preprint no: PM 95/05, hep-ph-9503234.
\bibitem{eichtinitial} E.~Eichten and B.~Hill, Phys. Lett. {\bf B234} 511 (1990).
\bibitem{eichten} A.~Duncan et al, Phys.~Rev.~{\bf D51}  5101 (1995).
\bibitem{bethlep} B.~A.~Thacker and P.~Lepage, Phys. Rev {\bf D43} 196 (1991).
\bibitem{biglep} P.~Lepage et al, Phys. Rev. {\bf D 46} 4052 (1992).
\bibitem{hash} S.~Hashimoto, Phys.~Rev.~{\bf D50}  4639 (1994).
\bibitem{cdavies} C.~T~H.~Davies, Nucl. Phys. {\bf B34} (Proc.~Suppl.)  437 (1994)
\bibitem{arifaspec} A.~Ali Khan et al, Glasgow preprint no: GUTPA/95/12/1, SCRI preprint no: FSU-SCRI-95-121, Ohio preprint no: OHSTPY-HEP-T-95-026, hep-lat-9512025. To be published in Phys.~Rev.~{\bf D53}.
\bibitem{spec} S.~Collins et al, SCRI preprint no: FSU-SCRI-96-13, Glasgow preprint no: GUTPA/96/2/8, Ohio preprint no: OHSTPY-HEP-T-96-002, Edinburgh preprint no: Edinburgh 96/1, hep-lat-9602028. Submitted to Phys.~Rev.~{\bf D}.
Note that the fit parameters for the multi-exponential fits differ
slightly~($\ltaeq 1\sigma$) from the results presented in this
reference. In that analysis a bootstrap procedure was used in
calculating the covariance matrix.  It was found, however, that the
advantages of this procedure did not outweigh the computational effort
involved and it is not implemented here.
\bibitem{hemcgc} K.~M.~Bitar et al, Phys.~Rev.~{\bf D46}  2169 (1992).
\bibitem{nrqcdups} C.~T.~H.~Davies et al, Phys.~Rev.~D~{\bf 50}, 6963 (1994).
\bibitem{clovernrqcd} S.~Collins et al, in preparation.
\bibitem{nrqcdrenorm} C.~Morningstar and J.~Shigemitsu, work in progress.
\bibitem{eichten} E.~Eichten and B.~Hill, Phys.~Lett.~{\bf B240}  193 (1990).
\bibitem{sach1} M.~Crisafulli et al, Nucl.~Phys.~{\bf B457} 594 (1995).
\bibitem{neubert} M.~Neubert, Phys.~Rev.~{\bf D46}  1076 (1992).
\bibitem{ukqcd} UKQCD Collaboration R.~Baxter et al,  Phys. Rev.~{\bf D49}
1594 (1994).
\bibitem{milc} C.~Bernard et al, Phys.~ReV.~{\bf D49} (1994) 2536. 
\bibitem{fermilab} T.~Onogi and J.~N.~Simone, Nucl.~Phys.~{\bf B} (Proc.~Suppl.) {\bf 42}  434 (1995).
\bibitem{booth} M.~J.~Booth, Florida preprint no: UFIFT-HEP-94-10, hep-ph-9412228.
\bibitem{arifadecay} A.~Ali Khan et al, in preparation.
\end{thebibliography}
\end{document}